\documentclass[iop,apj,tighten,numberedappendix,twocolappendix]{emulateapj}
\pdfoutput=1 
\usepackage{apjfonts} 
\usepackage{longtable,array}
\usepackage{amsmath,amstext}
\usepackage[breaklinks,colorlinks,citecolor=blue,linkcolor=magenta]{hyperref} 
\usepackage[all]{hypcap} 

\shorttitle{Gamma-Ray Bursts with RINGO2}
\shortauthors{Steele et al.}

\begin{document}
 
\title{Polarimetry and Photometry of Gamma-Ray Bursts with RINGO2}
\author{
I. A. Steele\altaffilmark{1},
D.~Kopa\v c\altaffilmark{1,2}, 
D. M. Arnold \altaffilmark{1},
R. J. Smith\altaffilmark{1},
S. Kobayashi\altaffilmark{1}
H. E. Jermak\altaffilmark{1,3},\\
C. G. Mundell\altaffilmark{4},
A. Gomboc\altaffilmark{5},
C. Guidorzi\altaffilmark{6}, 
A. Melandri\altaffilmark{7} and
J. Japelj\altaffilmark{8}
}
\affil{$^1$Astrophysics Research Institute, Liverpool John Moores University, Liverpool, L3 5RF, UK}
\affil{$^2$Faculty of Mathematics and Physics, University of Ljubljana, Jadranska 19, Ljubljana 1000, Slovenia}
\affil{$^3$Department of Physics, Lancaster University, Bailrigg Campus, Lancaster, LA1 4YW, UK}
\affil{$^4$Department of Physics, University of Bath, Claverton Down, Bath, BA2 7AY, UK}
\affil{$^5$Centre for Astrophysics and Cosmology, University of Nova Gorica, Vipavska 11c, Ajdov\v s\v cina 5270, Slovenia } 
\affil{$^6$Department of Physics and Earth Science, University of Ferrara, Italy}
\affil{$^7$INAF-Osservatorio Astronomico di Brera, via E. Bianchi 36, I-23807 Merate (LC), Italy}
\affil{$^8$Anton Pannekoek Institute for Astronomy, University of Amsterdam, Science Park 904, 1098 XH Amsterdam, The Netherlands}

\begin{abstract}
\par We present a catalog of early-time ($\sim10^2-10^4$s) photometry and polarimetry of all Gamma-Ray Burst (GRB) optical afterglows observed with RINGO2 imaging polarimeter on the Liverpool Telescope.  For the $19$ optical afterglows observed, the following $9$ were bright enough to perform photometry and attempt polarimetry: GRB\,100805A, GRB\,101112A, GRB\,110205A, GRB\,110726A, GRB\,120119A, GRB\,120308A, GRB\,120311A, 
GRB\,120326A and GRB\,120327A. 
We present multi-wavelength light curves for these 9 GRBs,
together with estimates of their optical polarization degrees and/or limits. We carry out a thorough investigation of detection probabilities, instrumental properties and systematics. Using two independent methods, we confirm previous reports of significant polarization in GRB 110205A and 120308A, and report new detection of $P=6^{+3}_{-2}$\% in GRB101112A. We discuss the results for the sample in the context of the reverse and forward shock afterglow scenario, and show that GRBs with detectable optical polarization at early time have clearly identifiable signatures of reverse-shock emission in their optical light curves. This supports the idea that GRB ejecta contain large-scale magnetic fields and highlights the importance of rapid-response polarimetry.
\end{abstract}

\maketitle

\section{Introduction}

Almost half a century since the discovery of Gamma-Ray Bursts (GRBs), these cosmic  explosions remain puzzling, particularly regarding the origin and role of magnetic  fields in driving the explosion \citep{granot}. Relativistic outflow associated  with GRB events is conventionally assumed to be a baryonic jet, producing synchrotron emission with tangled magnetic fields generated locally by instabilities in shocks \citep{piran1999,zhang2004}. However, recent polarization observations indicate the existence of large-scale magnetic fields in the outflow \citep{SteeleNature2009,Yonetoku,mundell-nature}. The rotation of a black hole and an accretion disk (i.e., the standard GRB central engine) might cause a helical outgoing magnetohydrodynamic wave which accelerates material frozen into the field lines. In such magnetic models, the outflow is expected to be threaded with globally ordered magnetic fields \citep{Komissarov}.

Because of their cosmological distances, measurement of the degree of polarization (P) and the electric vector polarization angle (EVPA) of the light is the only direct probe of magnetic fields in GRB jets. Early polarimetric studies focused on the evolution of polarization around a jet break to give constraints on the collimation of a jet and the angular dependence of the energy distribution \citep{Sari1999,GhiselliniLazzati1999,Rossi2004}. Jet breaks are expected to happen at $\gtrsim 1$ day after GRB triggers and observed polarization degrees at such late times are rather low at only a few per cent \citep{covino1999,Wijers1999}. 

Since the late-time afterglow is emitted from shocked ambient medium (i.e., forward shock), rather than the original ejecta from the GRB central engine, it is insensitive to the jet acceleration process. The magnetic properties of the original ejecta can be examined only through the investigation of the prompt gamma-rays or reverse shock emission. This requires polarization measurements of GRB themselves or the early afterglow ($\lesssim $ 30 mins). For this purpose, RINGO and RINGO2 imaging polarimeters on the Liverpool Telescope (LT) were developed, with which we can measure the polarization of afterglow just a few minutes after a GRB trigger. Since synchrotron emission is expected to be linearly polarized, only linear polarization measurements will be discussed in this paper (see \cite{Wiersema2014} for a recent detection of circular polarization and \cite{Nava2016} for discussion of its implication). Linear polarization also can be produced by the inverse Compton scattering process \citep{Lazzati2004,Lin2017}.

In this paper we present the complete catalog of photometry and polarimetry of GRBs observed with the RINGO2 imaging polarimeter on LT. Of the 19 optical afterglows observed, 9 were bright enough to perform photometry and attempt polarimetry. Additional photometric measurements obtained with RATCam \citep{ratcam} on the same telescope are also presented. RINGO2 technical details, calibration and the data reduction process are discussed in Section~\ref{sect:ringo2}. In Section \ref{sect:observations} we list the GRBs observed during RINGO2 operation, and in Sections~\ref{sect:photo-results} and \ref{sect:pol-results} we present the photometry and polarimetry results of the sample. A discussion and interpretation follows in Section~\ref{sect:discussion}, and we summarize our conclusions in Section~\ref{sect:conclusion}.

\section{RINGO2}
\label{sect:ringo2}
\subsection{Telescope and Instrument Description}

LT is a 2.0 meter fully robotic telescope at the Observatorio del Roque de los Muchachos, La Palma \citep{steele04}.  It can host multiple instruments with a rapid change time ($<30$ seconds) and is optimized for time domain astrophysics, including the rapid automated followup of transient sources such as GRBs \citep{pasp-paper}.  

RINGO2 \citep{steele2010} was operational on LT from 2010 August 1 to 2012 October 26.  Re-imaging optics gave the instrument a field of view of $4\times4$ arcmin.   It used a polarizer rotating at $\sim1$-Hz to modulate the incoming beam from the telescope and a fast readout, low noise electron multiplying CCD (EMCCD) camera to sample the modulated image.  Readout of the camera was electronically synchronized to the polarizer angle such that exactly 8 images were obtained for a single rotation.  By analysis of the relative intensities of a source within the 8 images, the degree of polarization could be determined. 

All data from RINGO2 are pipeline reduced in the telescope's computer system to remove the standard instrumental signatures associated with CCD imaging.  This comprises dark subtraction and flat-field division.  Due to the short individual exposure times ($\sim$ 125 ms), an observation will comprise many repeated exposures at each of the eight rotor positions.  Each rotor position exposure is therefore combined in longer ($1-10$ minute) time bins to make 8 mean images (one per rotor position). 
A world coordinate system (WCS) fit is then added to the FITS headers and the mean images are transferred to the user for analysis.

\subsection{Extraction of Polarization Signal}
\label{sec:pol-extract}

To extract the polarization signal for an object from RINGO2 data it is necessary to measure the relative number of (sky subtracted) counts in each of the 8 mean images for that object.   Due to the field position dependent point spread function (PSF) caused by the RINGO2 re-imaging optics, PSF fitting was not appropriate for this measurement.   Instead we used aperture photometry with an aperture size of 3.5 arcsec diameter.  This value is the mean location of the maximum in a signal-to-noise ratio versus aperture size plot for multiple observations of 10 objects in the  field of the polarimetric standard HD212311. The objects had apparent magnitudes in the range 8 to 17.     
Extraction of the counts for every source on every image was automated using {\em SExtractor} \citep{sextractor} with local sky subtraction.  More detailed
descriptions of these procedures are presented in \cite{jermak-mnras, jermak-phd, arnold}.

\begin{figure}[!h]
\begin{center}
\includegraphics[angle=90, width=8.5cm]{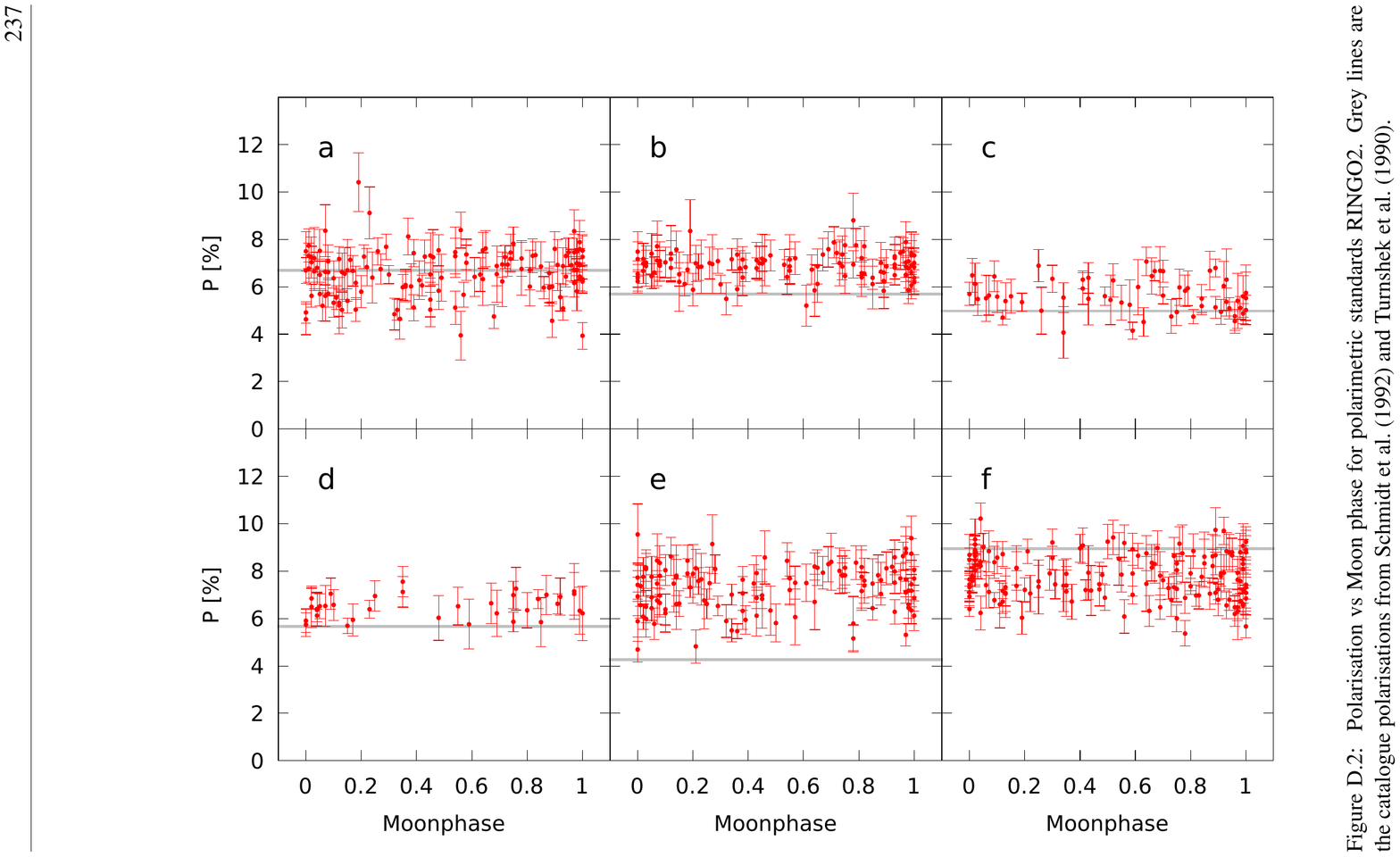} 
\caption{\label{fig:moon}  Observed polarization versus Lunar phase for the polarized standard stars
(a) BD +49 389,  (b) BD+64 106, (c) HD 155528,  (d) Hiltner 960,  (e) BD +25 727, (f) VI Cyg \#12.
The horizontal gray lines are catalog polarizations from \cite{schmidt} and \cite{turnshek}}
\end{center}
\end{figure}

For every object in the 8 image set, the measured Stokes $q_m$ and $u_m$ parameters and associated errors (based on the photon statistics) are calculated from the sky subtracted counts following the prescription presented by \cite{clarke}.  These results are stored in a MySQL database along with other FITS header data such as observation date and telescope and environmental information. This allowed checks to be made for any trends with quantities such as lunar phase (Figure \ref{fig:moon}) in the data.  No such trends were found (see \citet{arnold} for more details).

\begin{figure}[!h]
\begin{center}
\includegraphics[angle=0, width=8.5cm]{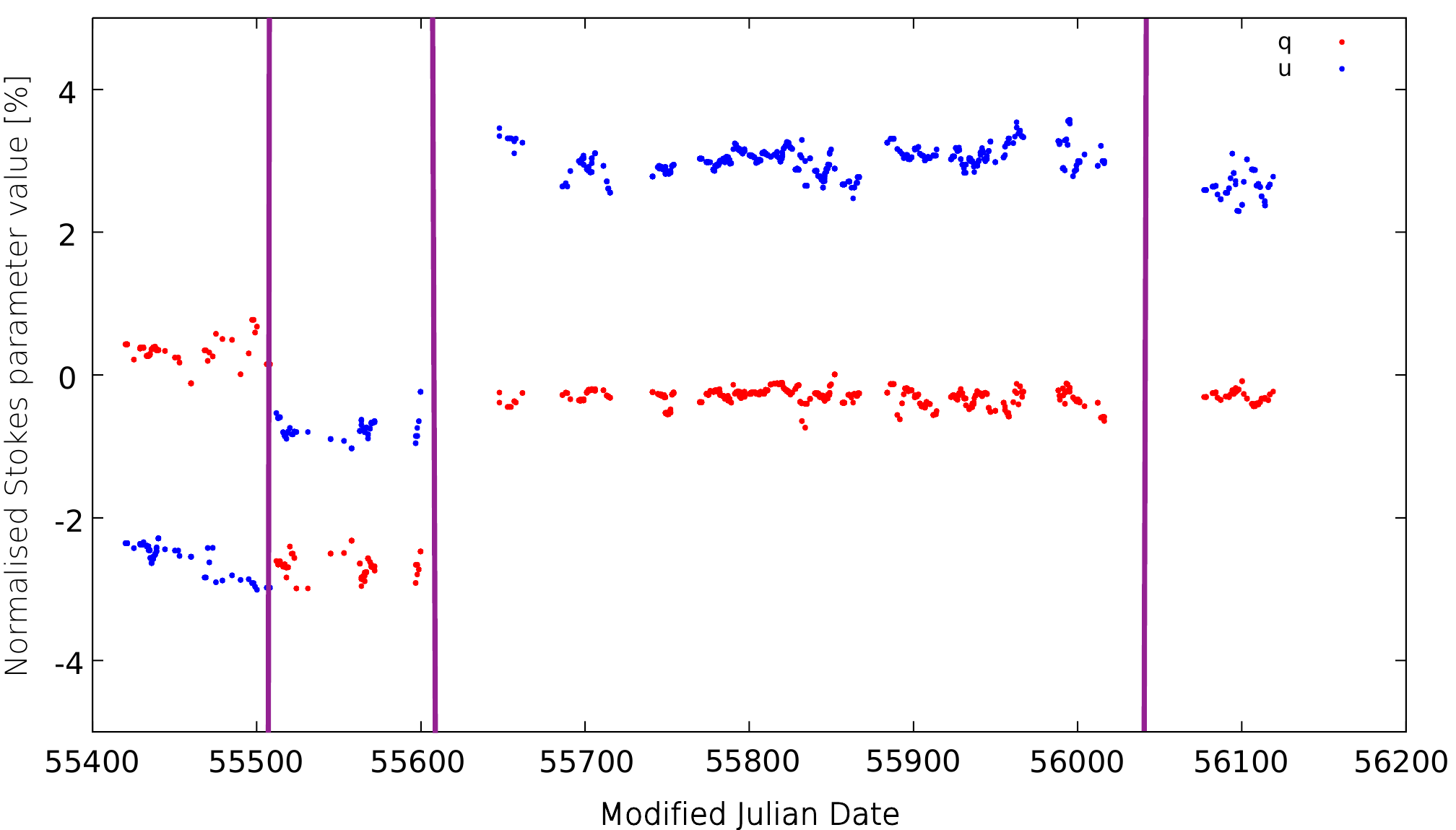} 
\caption{\label{fig:npol}  Time evolution of the measured RINGO2 instrumental polarization zeropoints.  Blue points indicate $u_0$ and red points $q_0$.  Times of instrument hardware changes are indicated by vertical lines.}
\end{center}
\end{figure}

Conversion of the measured Stokes $q_m$ and $u_m$ parameters to degree of polarization ($P$) and electric vector polarization angle (EVPA) is carried out via the standard equations

\begin{equation}
q=q_m-q_0
\end{equation}

\begin{equation}
u=u_m-u_0
\end{equation}

\begin{equation}
P=\frac{\sqrt{q^2 + u^2}}{D}
\end{equation}

\begin{equation}
{\rm EVPA} = \frac{1}{2}\arctan{\frac{u}{q}} + {\rm SKYPA} + {\rm K}
\end{equation}

\noindent where $q_0$ and $u_0$ are measures of the instrumental polarization.  
$q_0$ and $u_0$ were determined using our observations of zero polarized standard stars \citep{schmidt}. 
Figure \ref{fig:npol} shows the results of this analysis as a function of time. The final derived quantities based on combining data from all of the standards is presented in Table \ref{tab:ringo-cal}. 
Step changes in these quantities are associated with instrument servicing activity.  Apart from that they remain constant.

Though it rotates between $q_0$ and $u_0$ with hardware servicing activities, the mean measured instrumental polarization over the entire RINGO2 lifetime was a nearly constant 2.9\%, with a standard deviation of 0.4\%.  This (relatively high) instrumental polarization is mainly caused by the instrument being fed from a 45$^\circ$ reflecting mirror in the telescope.  

The quantity $D$ is a measure of the instrumental depolarization caused by the imperfect contrast ratio of the polarizer.  It was calibrated by observations of polarized standard stars \citep{schmidt,turnshek} over the entire observation period and determined to be $0.76\pm0.01$.  
Using this value to correct the measured instrumental polarization gives a value of $3.8\%$ in line with the expectation for a 45$^\circ$ reflecting mirror \citep{cox}

The quantity SKYPA is the telescope Cassegrain axis sky 
position angle (measured East of North) and $K$ a calibration offset to that position angle that combines the angles between the orientation of the polarizer, the telescope focal plane and the trigger position of the angle measuring sensor.  $K$ was determined using our polarized standard star observations and was found (Table \ref{tab:ringo-cal}) to be stable within each observation period to within 4$^\circ$ (standard deviation).

An analysis of the position dependence of polarization in the instrument derived from observations of the twilight sky is presented in the supplementary material of \cite{mundell-nature}. This shows that this effect is $<1.5$\% of the measured polarization (so for example on a 10\% polarized source it would introduce a maximum error of 0.15\%).

Since $q$ and $u$ are constructed by linear combinations of count values that are subject to Poisson counting statistics, their error distributions will be normally distributed (symmetrical)  and can be calculated by
standard error propagation theory \citep{clarke}.  However, for $P$ and EVPA the process is more complex. In particular $P$, being a quantity that is always positive (Equation 3) will have a
Rayleigh distribution \citep{rayleigh}.  This means it will have an asymmetric distribution of errors.  The value of $P$ itself
will also therefore suffer from a polarization bias where noise in $q$ and $u$ will generate a false increase in the $P$ value \citep{simmons}.  A similar problem also affects EVPA measurements \citep{nag93}.  These problems must be particularly addressed at low values of $P$ where the error distribution becomes increasingly asymmetric.  They are taken into account in the analysis of our GRB results presented in  \ref{sect:pol-results}.

\renewcommand{\arraystretch}{1.2}
\begin{table*}[]
\scriptsize
\begin{center}
\caption{Mean Stokes $q$ and $u$ zeropoints and EVPA zeropoint $K$ for RINGO2 periods of operation.  $pm$ errors are the standard error on the mean, whereas sd() quantities are the standard deviation of the sample.} 
\label{tab:ringo-cal}
\begin{tabular}{llllllll}
\hline
\hline
MJD Range & Date Range & $q_m$ & sd($q_m$) & $u_m$ & sd($u_m$) & K & sd($K$) \\
\hline
55418--55510& $20100810 - 20101110$ & $+0.0030\pm0.0006$  & 0.0031 & $-0.0250\pm0.0006 $& 0.0041& 126$^\circ$  & 4$^\circ$\\
55511--55607& $20101111 - 20110215$ & $-0.0261\pm0.0005$ &  0.0047    & $-0.0074\pm0.0005$   &   0.0031         &  171$^\circ$ & $3^\circ$ \\
55640--56045& $20110320 - 20120428$ & $-0.0030\pm0.0002$ & 0.0025 & $+0.0297\pm0.0003 $& 0.0036 & 41$^\circ$ & 3$^\circ$\\
56045--56226& $20120428 - 20121026$ & $-0.0031\pm0.0004$ &0.0017 & $+0.0264\pm0.0005 $& 0.0041 & 42$^\circ$ & 2$^\circ$\\
\hline
\end{tabular}
\end{center}
\end{table*}


\subsection{Photometric reduction and calibration}

In addition to the RINGO2 observations, optical band photometry of each burst was carried out  using the RATCam CCD imaging camera in intervals between and after the RINGO2 observations. These photometric observations were typically using either $g'r'i'$ or $r'i'z'$ filter sequences and provide multi-color light curves that cover a longer time baseline than the RINGO2 observations alone. Conventional circular aperture photometry was performed with sky flux determined locally for each source from an annular aperture surrounding the target. Zero points were derived from RATCam observations of SDSS secondary standards \citep{sloan} taken on the same night. Instrumental zero points in each filter were obtained as an average for all the SDSS standards available, which amounted to between two and five different stars per night. Rather than apply this zero point directly to photometry of the GRB afterglow, a zero point was established for each GRB frame as the average for several sources detected in that frame with comparable brightness to the GRB afterglow. The number of available stars varied between two and seven. Any sources which showed statistically significant variation during the period of observation, whether that be genuine variability of simply poor data, were rejected before deriving the instrumental zero point for that image. The optical transient magnitude was finally obtained by aperture photometry relative to that ensemble average of between two and five field stars.

No color corrections have been included because detailed transformations between RATCam and the SDSS calibration telescope are not available. However, the RATCam filters are sufficiently close to the SDSS passbands that errors are expected to be substantially smaller than the typical statistical photon counting errors in our observations. The RATCam observers' documentation cites color corrections of less than 0.05($r'-i'$), implying less than 0.025\,mag for sources of typical stellar colors.

\begin{figure}[!h]
\begin{center}
\includegraphics[angle=270, width=1\linewidth]{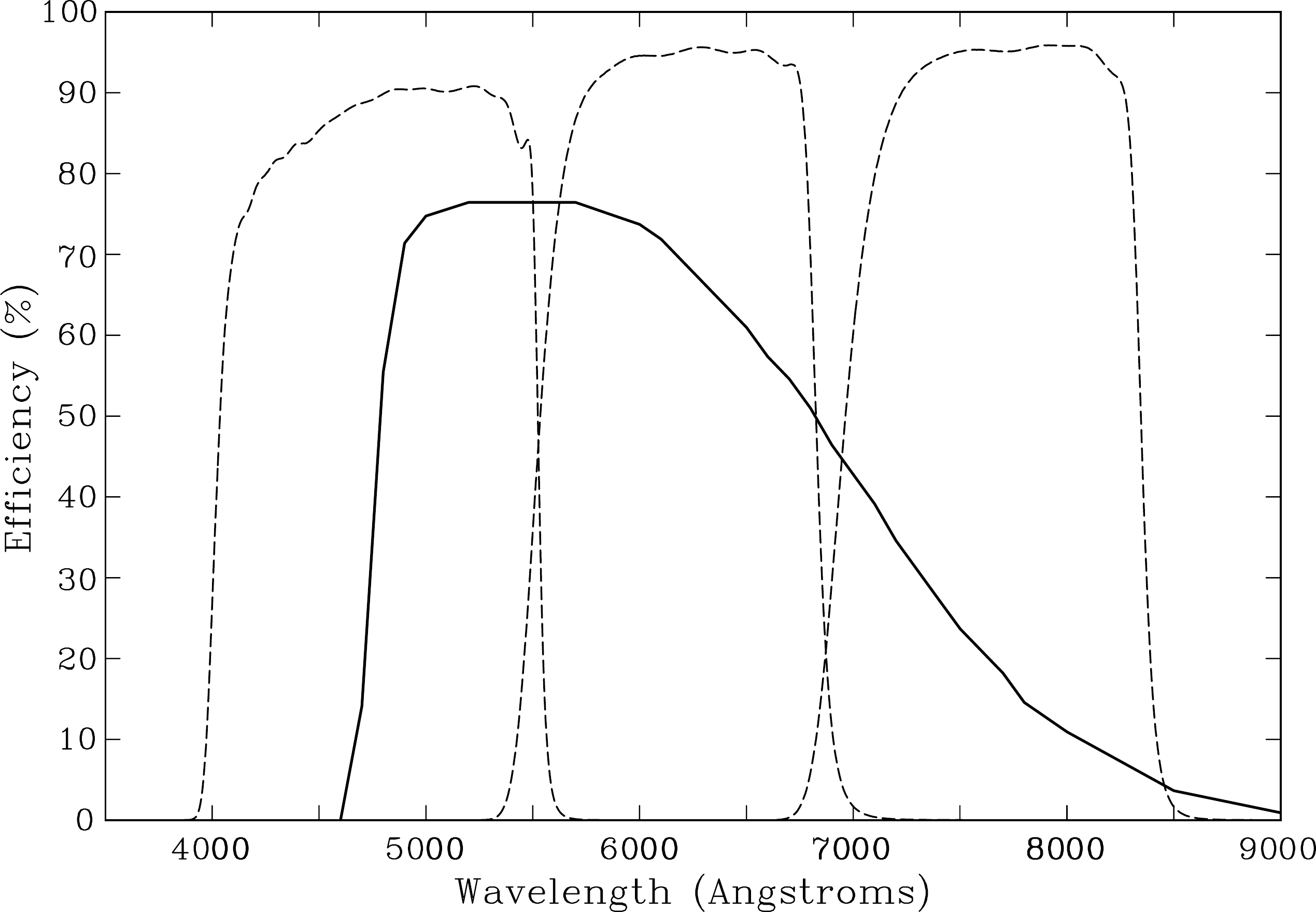} 
\caption{\label{fig:ringo-filter} RINGO2 filter throughput (solid line) compared to SDSS-$g',r',i'$ filters (dashed lines from left to right)}
\end{center}
\end{figure}

\begin{figure*}[t]
\begin{center}
\includegraphics[width=1\linewidth]{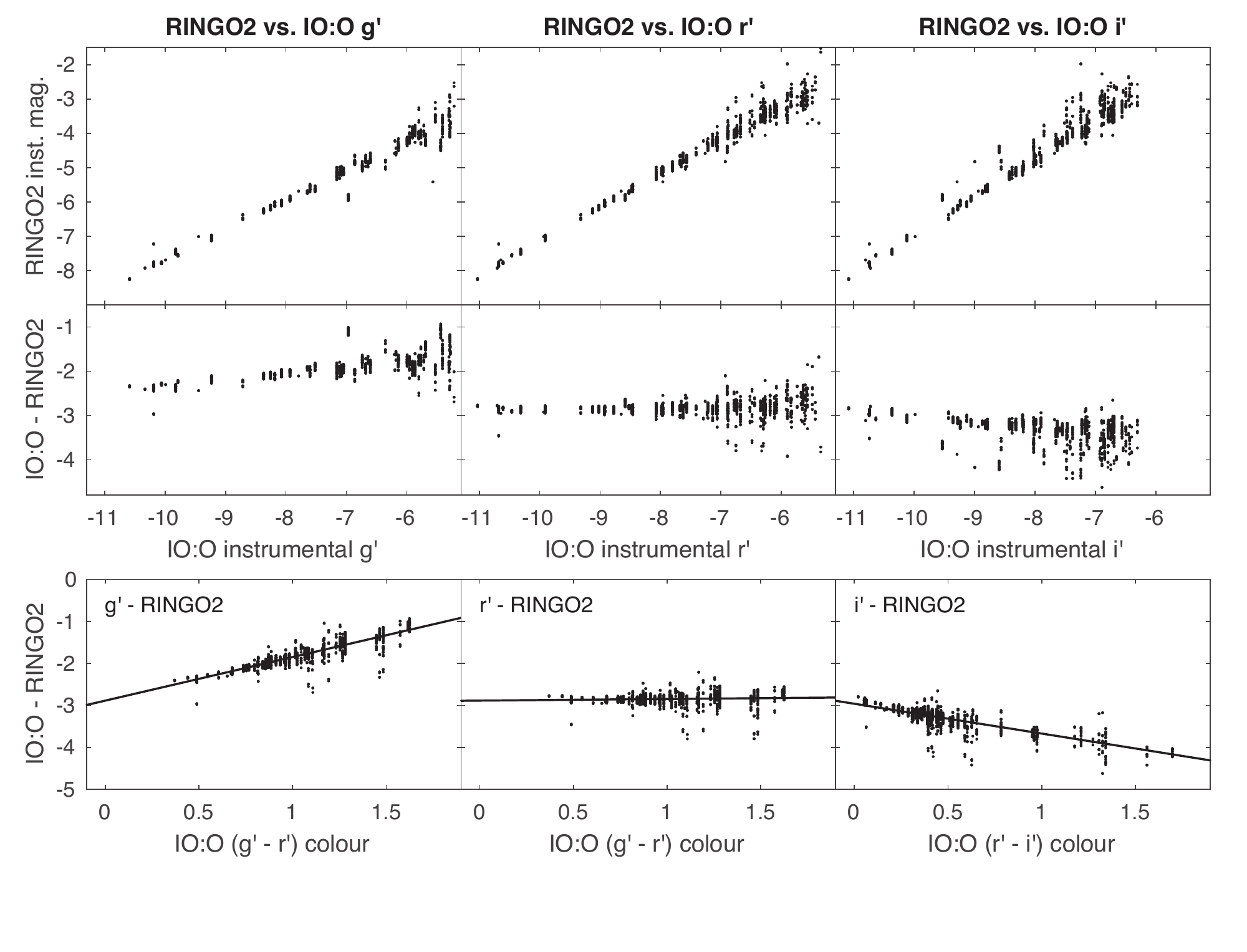}
\caption{\label{fig:ringo-colour} Aperture photometry of all stars in the field surrounding star BD+32 3739 (HD 331891). RINGO2 data are taken from observations of this standard star on all photometric nights between 2012 June 3 and 2012 October 26. The eight RINGO2 images at each epoch are co-added to create an unpolarized image. The comparison IO:O data are a single epoch, obtained on the night of 2013 September 4. All numbers are simple instrumental magnitudes calculated as $-2.5\,{\rm log}_{10}(counts)$. The left panels compare RINGO2 to IO:O$+g'$ filter, the central column is RINGO2 and IO:O$+r'$ and the right panels show IO:O$+i'$. The top row directly compares the instrumental magnitudes from the two instruments. The middle row plots the magnitude difference between RINGO2 and various IO:O filters, effectively the zero point difference between the instruments, which is shown to be independent of magnitude for filter $r'$. The bottom row derives zero point color transformations between the RINGO2 filter and the various SDSS-type filters. Again $r'$ is seen to be a good match to RINGO2 without applying any color correction. The clumpy distribution of points is caused by a single IO:O observation of each star being compared to 54 epochs of RINGO2 observations.}
\end{center}
\end{figure*}

Summing the eight polarised images in a RINGO2 observation provides unpolarized photometry.  The wavelength range is determined by a custom filter comprised of 3mm Schott GG475 cemented to 2mm Schott KG3 which gives an approximate wavelength range of $4750 - 7100$\,{\AA} ({\bf 2350\,\AA} FWHM).  The filter bandpass is shown in Figure \ref{fig:ringo-filter} where it can be compared to the SDSS filter band passes \citep{sloan}.  For each afterglow we therefore have both multi-filter RATCam and single-filter RINGO2 imaging. The primary objective here is to transform the RINGO2 data onto a similar reference frame as the RATCam data and allow direct comparison in a single light curve.

As part of its routine calibration program, RINGO2 observed zero and non-zero polarized standard objects \citep{schmidt} several times per night throughout its period of operation.  To characterize the non-standard filter, observations have also been made of the same fields with the CCD imager IO:O \citep{steele14} which replaced the decommissioned RATCam in June 2012. IO:O has a filter wheel and full suite of SDSS-type filters. These are polarimetric not photometric standards so the fields do not contain established references intended to develop an absolute photometric calibration.  However we can use them to compare raw instrumental magnitudes of the many field stars in the various SDSS filters with RINGO2 magnitudes. Figure \ref{fig:ringo-colour} demonstrates that SDSS $r'$ provides an excellent match to RINGO2's natural photometric system.  We therefore use SDSS $r'$ as the basis for the relative photometric calibration of our GRB afterglow lightcurves with no need to apply color corrections.

Having selected SDSS r' as the best comparison reference, photometry was extracted from the RINGO2 frames following the same procedures as the RATCam data described above.

For both RATCam and RINGO2, the magnitudes of the optical transient (OT) in various bands were corrected for Galactic extinction using maps from \cite{extinction}\footnote{http://ned.ipac.caltech.edu/forms/calculator.html}. Finally we converted to flux densities using flux zero points provided in \cite{fuk}.

\section{Observations}
\label{sect:observations}
Between 2010 and 2012, $19$ optical afterglows were observed with RINGO2 polarimeter. Table~\ref{tab:sample_complete} shows observational properties of the complete sample, the time of the RINGO2 observations and the mid time optical ($r'$ equivalent band) magnitude of the source. In most cases, the LT and RINGO2 response time was 2--3 minutes, but only one event (GRB\,101112A) was brighter than $\sim 16$th magnitude during these observations. Among $19$ afterglows, $10$ were too faint during the time of RINGO2 observations to perform photometry and thus polarimetry. For the remaining $9$ we were able to perform both photometry and attempt polarimetry. The results for GRB\,120308A \citep{mundell-nature} and a preliminary analysis of GRB\,110205A \citep{cucchiara} have already been presented separately.  

Individual details of the RINGO2 and RATCam photometric reduction for each GRB are provided in Section \ref{sect:photo-results}.  The RINGO2 polarimetric results are presented in Section \ref{sect:pol-results}.

\renewcommand{\arraystretch}{1.2}
\begin{table}[]
\scriptsize
\begin{center}
\caption{The complete sample of $19$ GRB afterglows observed with RINGO2. Photometric errors quoted are the statistical error on the observation. In particular those calibrated with respect to USNO-B1 catalogs (GRB\,110726A, GRB\,120326A) show larger systematic errors
(typically 0.3\,mag).}
\begin{tabular}{llll}
\hline
\hline
GRB & $t-T_0$ [$\mathrm{s}$] & $r'$ Mag. at mid time & GCN Reference \\
\hline
100802A & $116 - 295$ & $18.79 \pm 0.64$  \\
100805A & $140 - 320$ & $17.29 \pm 0.13$  \\
 & $1020 - 1198$ & $18.76 \pm 0.57$ \\
101112A & $176 - 355$ & $15.77 \pm 0.03$ \\
 & $715 - 893$ & $16.61 \pm 0.05$ \\
110106B & $697 - 875$ & $>22.5$  at 2620s & \cite{gcn11537} \\
110205A & $422 - 722$ & $16.92 \pm 0.68$ \\
 & $3026 - 3506$ & $16.37 \pm 0.07$ \\
110402A & $214 - 813$ & $\sim 20.8$ at 1680s & \cite{gcn11858} \\
110520A & $142 - 741$ & $>19.0$ at 215s & \cite{gcn12022}\\
 & $1081 - 1259$ & $>19.7$ at 1879s & \cite{gcn12024}\\
110726A & $191 - 783$ & $17.99 \pm 0.11$  \\
 & $4582 - 5180$ & ... \\
120119A & $194 - 793$ & $17.65 \pm 0.04$  \\
120305A & $154 - 752$ & $> 21.3$ at 840s & \cite{gcn13006}  \\
120308A & $240 - 838$ & $16.51 \pm 0.03$ \\
120311A & $181 - 779$ & $18.41 \pm 0.18$  \\
 & $3818 - 4416$ & ... \\
120324A & $183 - 781$ & $>20.3$ at 840s & \cite{gcn13092} \\
120326A & $216 - 872$ & $18.88 \pm 0.14$  \\
120327A & $1664 - 2263$ & $16.66 \pm 0.03$ \\
 & $2605 - 2784$ & $17.11 \pm 0.05$ \\
120514A & $556 - 1155$ & $> 18.1$ at 721s & \cite{gcn13290} \\
120521C & $777 - 1375$ & $> 21.5$ at 2100s & \cite{gcn13320} \\
120711B & $157 - 755$ & $>21.7$ at 42469s & \cite{gcn13465} \\
 & $1249 - 1847$ & ... \\
 & $2403 - 3001$ & ... \\
120805A & $215 - 813$ & $\sim 20.9$ at 960s & \cite{gcn13651} \\
\hline
\end{tabular}
\end{center}

{\bf Notes.} All bursts were initially detected by {\em Swift} apart from GRB101112A which was detected by {\em INTEGRAL}.  The second column shows the time of RINGO2 observations with respect to the Gamma Ray trigger and  the third column the $r'$ band optical magnitude at the mid-time of the RINGO2 epoch.    Where there is no $r'$ magnitude with error reported, the source was too faint to be observed with RINGO2 and a limit or measurement from the GCN report nearest in time is given.
\label{tab:sample_complete}
\end{table}

\section{Results: Photometry}
\label{sect:photo-results}

\begin{figure*}[!h]
\begin{center}
\includegraphics[width=18cm,angle=0]{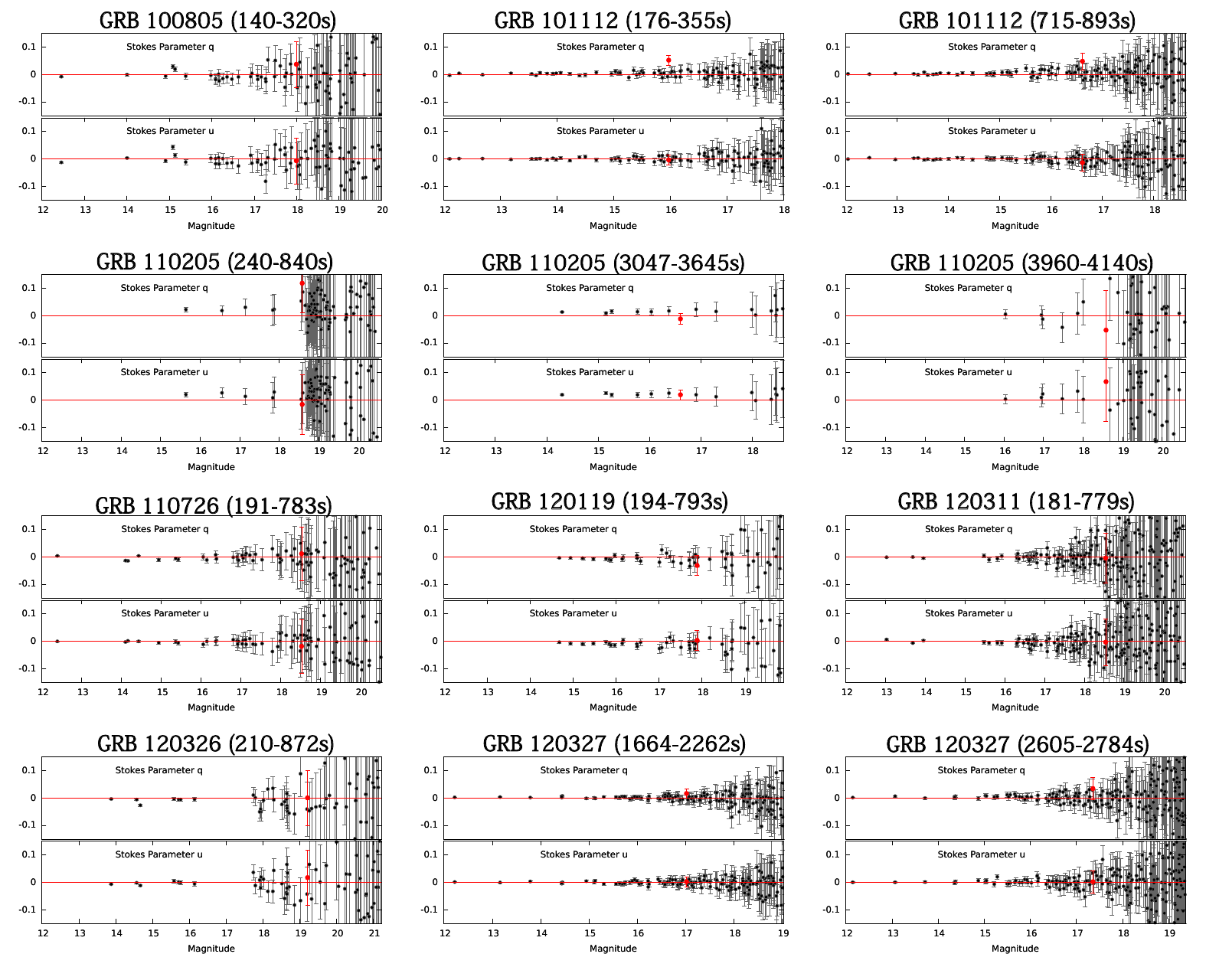} 
\caption{\label{fig:qu-mag} Stokes $q$ and $u$ parameters for the GRB sample (excluding GRB\,120308A) as a function of apparent magnitude.  In each plot the red point indicates the GRB and the
black points are the other objects in the same frame.  The $q$ and $u$ values have been corrected for instrumental polarization using the measured instrumental zeropoints.  The data have {\em not} been corrected for instrumental depolarization. }
\end{center}
\end{figure*}

\begin{figure*}[!h]
\begin{center}
\includegraphics[width=18cm,angle=0]{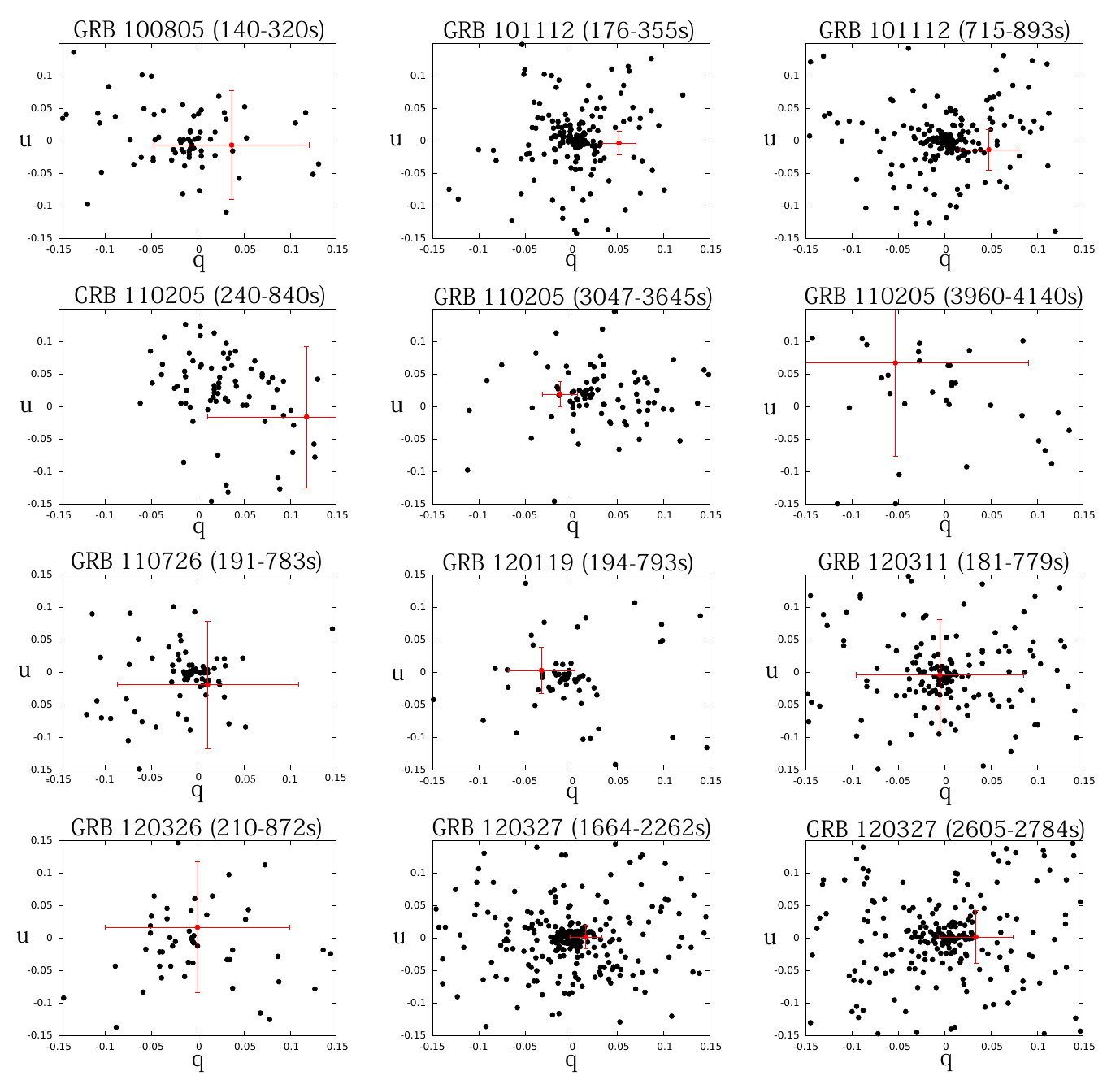} 
\caption{\label{fig:stokes} Stokes $q$ versus $u$ parameters for the GRB sample (excluding GRB\,120308A).  The red point (with error bars) indicates the GRB and the
black points (without error bars) are the other objects in the same frame.  The $q$ and $u$ values have been corrected for instrumental polarization using the measured instrumental zeropoints.  The data have {\em not} been corrected for instrumental depolarization. }
\end{center}
\end{figure*}

\begin{figure*}[!h]
\begin{center}
\includegraphics[width=18cm, angle=0]{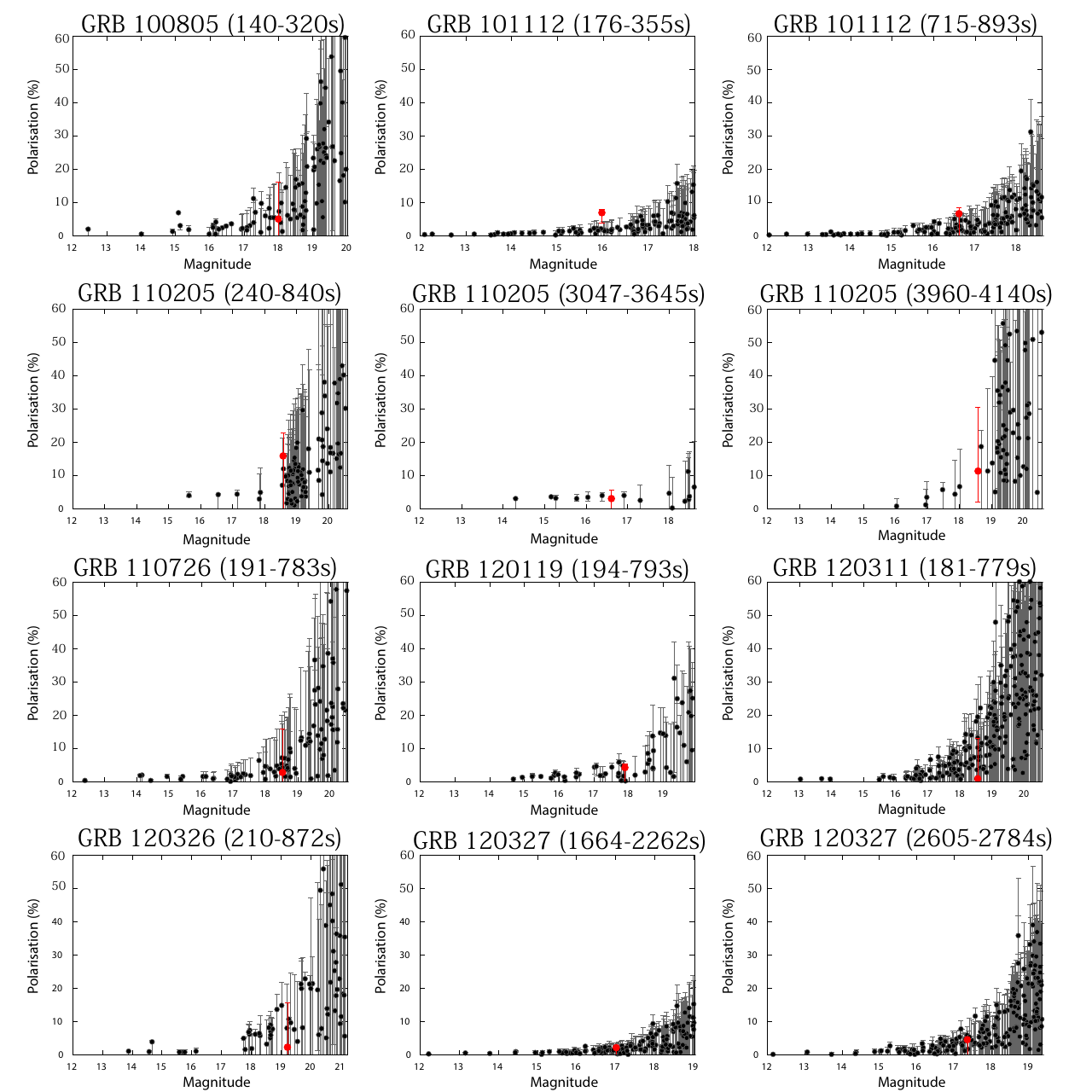} 
\caption{\label{fig:trumpet} Observed polarization for GRBs (red points) and the other objects in the same frame (black points) as a function of magnitude. The magnitude is derived directly from the count rate of each object assuming a constant zeropoint and is therefore not corrected for variations in sky transparency between different frames.  The data are corrected for instrumental polarization and depolarization.
The error bars are calculated from a Monte Carlo simulation based on simulating a range of $P$ from 0.01 to 70.00\% for each object and then generating a distribution of q and u values and hence a distribution of simulated possible measured $P$ values corresponding to the input distribution.  If the observed $P$ lies in the 14th-86th percentile (corresponding to 1$\sigma$ for a Gaussian distribution) then it is flagged as a possible true $P$ value.  The highest and lowest possible $P$ values therefore give the error bar. As the count rate decreases both the polarimetric error and polarization value increase.  The apparent increase in polarization value is the effect of polarization bias where noise is transformed into signal via the polarization equation (3).}
\end{center}
\end{figure*}

\begin{figure*}[!h]
\begin{center}
\includegraphics[width=18cm,angle=0]{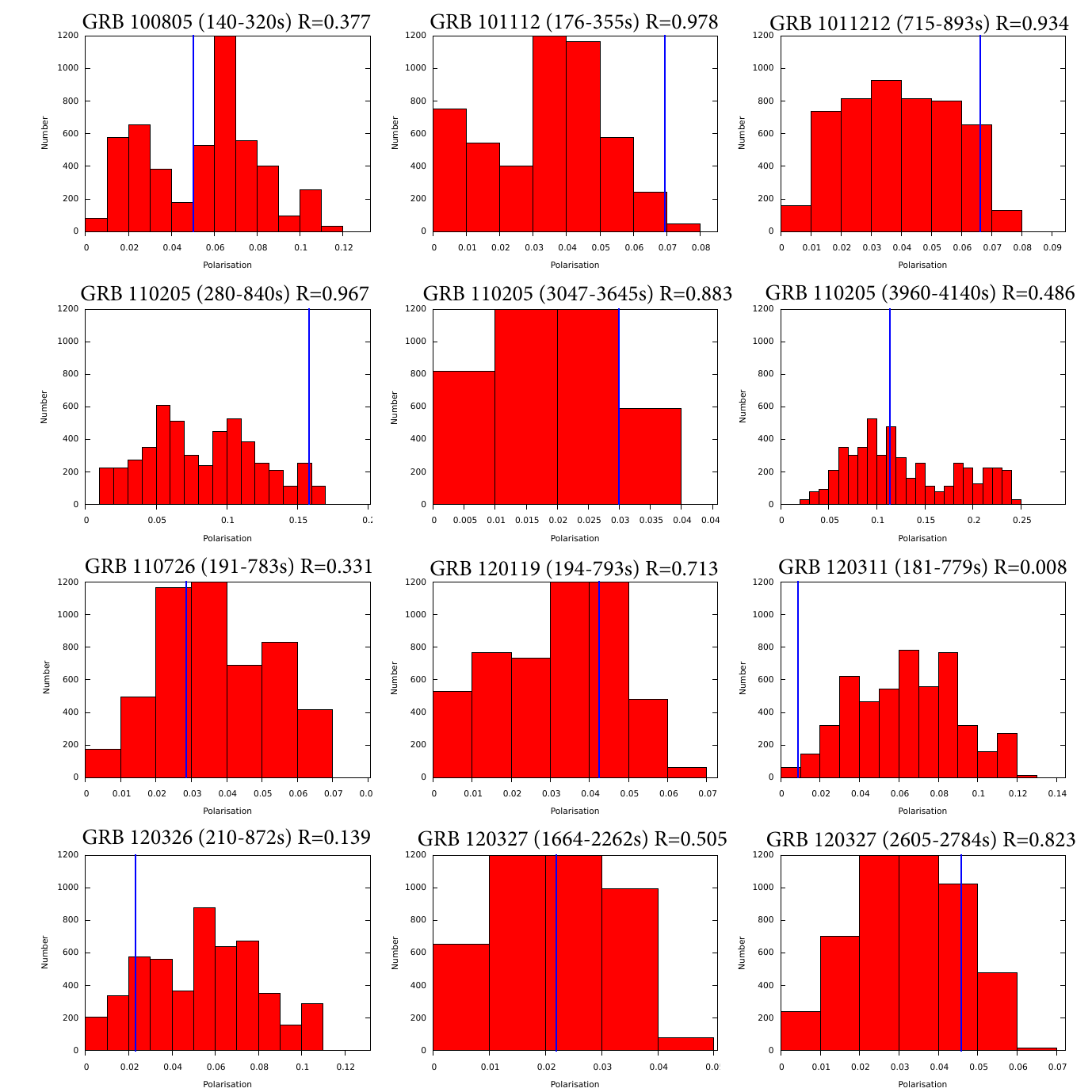} 
\caption{\label{fig:hist} Distribution of possible polarization values for each GRB. The histograms are constructed using all possible permutations of the 8 flux values measured for that GRB and have equal bin size $P=1$\%.  The data are corrected for both instrumental polarization and depolarization. The vertical (blue) line indicates the measured value for the GRB.  The normalized rank (R) of the GRB within the distribution is indicated.  For sources with R$<0.9$ we conclude the measured flux values are consistent with zero polarization and use the maximum permutated value to derive an upper limit.}
\end{center}
\end{figure*}

In the following subsections we summarize photometric reduction and calibration details for $9$ GRBs from RINGO2 sample, to obtain the complete light curves which are tabulated in the Appendix. All data were processed according the general procedures described above and only particular features of individual data sets are described here. In addition, high-energy properties from literature are also provided for all $9$ GRBs in the sample.

\subsection{GRB\,100805A}
RATCam observations  were  obtained in SDSS $g'$, $r'$ and $i'$. Final OT photometry 
is quoted with respect to a zero point established by the average of four field stars. Galactic extinctions applied were $A_{g'} = 0.61$, $A_{r'} = 0.42$, $A_{i'} = 0.31$.

The gamma-ray duration for this GRB in the $15-350\,\mathrm{keV}$ band is $\mathrm{T_{90}} = 16.7 \pm 3.1 \,\mathrm{s}$ \citep{lien16}, while the photometric redshift obtained from the UVOT data is $z \approx 1.3$ \citep{gcn11062}.

\subsection{GRB\,101112A}
The optical afterglow was discovered by \cite{gcn11397}. RATCam data are available in SDSS $g'$, $r'$ and $i'$ filters. Three field stars were averaged to give the frame reference in the OT images.
Galactic extinction corrections are $A_{g'} = 0.50$, $A_{r'} = 0.35$, $A_{i'} = 0.26$.

The gamma-ray duration for this GRB in the $50-300\,\mathrm{keV}$ band is $\mathrm{T_{90}} = 9.2 \,\mathrm{s}$ \citep{gcn11403}, while the upper limit on the photometric redshift (given the detection in $g'$ band) is $z \lesssim 3.5$.

\subsection{GRB\,110205A}
RATCam data in SDSS $g'$, $r'$ and $i'$, and SWIFT UVOT u, b and v magnitudes were all obtained from \cite{cucchiara}. Photometry from the LT RATCam data was re-checked following the same procedures as the other targets presented here
and found to be consistent with the published data. The additional UVOT data from \cite{cucchiara} allow better coverage across the optical peak. 
Galactic extinction corrections are $A_{V} = 0.04$, $A_{R} = 0.03$, $A_{g'} = 0.05$, $A_{r'} = 0.03$, $A_{i'} = 0.02$, $A_{u} = 0.06$, $A_{b} = 0.05$, $A_{v} = 0.04$. The SDSS magnitudes were converted to flux densities as previously described and the UVOT magnitudes using flux zero points provided in \cite{breeveld}.

The gamma-ray duration for this GRB in $15-350\,\mathrm{keV}$ band is $\mathrm{T_{90}} = 249 \pm 15 \,\mathrm{s}$ \citep{lien16}, while the redshift is $z=2.22$ \citep{gcn11638}.

\subsection{GRB\,110726A}

RATCam data in $g'$, $r'$ and $i'$ filters were obtained, but same night observations of photometric standards were not available. Instead, B2, R2 and I magnitudes from USNO-B1 were converted to SDSS $g'$, $r'$ and $i'$ magnitudes using \cite{jordi} and combined with aperture photometry of four nearby field stars to provide a zero point on each image. The errors tabulated in the Appendix are statistical estimates (consistent with our treatment of the other afterglows) that do not take into account that USNO-B1 has a typical, spatially varying, systematic photometric error ($1\sigma$) of $\sim 0.3$ magnitude. The Galactic extinction corrections were $A_{V} = 0.21$, $A_{R} = 0.17$, $A_{g'} = 0.26$, $A_{r'} = 0.18$, $A_{i'} = 0.13$.

The gamma-ray duration for this GRB in the $15-350\,\mathrm{keV}$ band is $\mathrm{T_{90}} = 5.2 \pm 1.1 \,\mathrm{s}$ \citep{lien16}, while the redshift range is $1.036 < z < 2.7$ \citep{gcn12202}.

\subsection{GRB\,120119A}

The RATCam observations used SDSS $r'$, $i'$ and $z'$ filters. Our photometry is found to be consistent with data published in \cite{morgan}. Additional PROMPT data in R and I filters from that paper were used to sample the light curve more densely at early and late times. Galactic extinction corrections are $A_{R} = 0.23$, $A_{I} = 0.16$, $A_{r'} = 0.25$, $A_{i'} = 0.18$, $A_{z'} = 0.14$.

The gamma-ray duration for this GRB in the $15-350\,\mathrm{keV}$ band is $\mathrm{T_{90}} = 68.0 \pm 7.1 \,\mathrm{s}$ \citep{lien16}, while the redshift is $z = 1.728$ \citep{gcn12865}.

\subsection{GRB\,120308A}
The RATcam observations used $r'$, $i'$ and $z'$ filters. Photometry was performed using the procedures outlined above and have already been published in \cite{mundell-nature}. As an extra cross-check, that paper also derived magnitudes using USNO-B1 magnitudes for which we transformed B2, R2 and I magnitudes to SDSS  $r'$, $i'$ and $z'$ magnitudes using \cite{jordi}. The results were consistent within $1\sigma$ error bars. The Galactic extinction maps give $A_{r'} = 0.09$, $A_{i'} = 0.07$, $A_{z'} = 0.05$.

The gamma-ray duration for this GRB in the $15-350\,\mathrm{keV}$ band is $\mathrm{T_{90}} = 61 \pm 17 \,\mathrm{s}$ \citep{lien16}, while the derived photometric redshift is $z \approx 2.2$ \citep{mundell-nature}.

\subsection{GRB\,120311A}
RATCam observations used $r'$, $i'$ and $z'$ filters and zero points were derived from the average of three field stars. Galactic extinction  values for the field are $A_{r'} = 0.31$, $A_{i'} = 0.23$, $A_{z'} = 0.17$.

The gamma-ray duration for this GRB in the $15-350\,\mathrm{keV}$ band is $\mathrm{T_{90}} = 3.5 \pm 0.8 \,\mathrm{s}$ \citep{lien16}, while the derived upper limit on the photometric redshift is $z \lesssim 3$ \citep{gcn13051}.

\subsection{GRB\,120326A}
RATCam data in $r'$, $i'$ and $z'$ filters were obtained from \cite{melandri}. Same night observations of photometric standards were not available so each frame's zero point was derived by averaging nearby field stars using USNO-B1 magnitudes as the reference. See the note regarding USNO-B1 errors for GRB\,110726A. The Galactic extinction corrections were $A_{V} = 0.14$, $A_{R} = 0.11$, $A_{r'} = 0.12$, $A_{i'} = 0.09$, $A_{z'} = 0.06$.

The gamma-ray duration for this GRB in the $15-350\,\mathrm{keV}$ band is $\mathrm{T_{90}} = 69.5 \pm 8.2 \,\mathrm{s}$ \citep{lien16}, while the redshift is $z = 1.798$ \citep{gcn13118}.

\subsection{GRB\,120327A}
RATCam observations were in $r'$, $i'$ and $z'$ filters and zero points derived from average of three field stars.  Galactic extinction correction for the field is $A_{r'} = 0.76$, $A_{i'} = 0.57$, $A_{z'} = 0.42$.

The gamma-ray duration for this GRB in the $15-350\,\mathrm{keV}$ band is $\mathrm{T_{90}} = 63.5 \pm 7.0 \,\mathrm{s}$ \citep{lien16}, while the redshift is $z = 2.81$ \citep{gcn13133}.

\section{Results: Polarimetry}
\label{sect:pol-results}

\renewcommand{\arraystretch}{1.2}
\begin{table*}[]\footnotesize
\begin{center}
\caption{Polarization results.}
\begin{tabular}{cccccccccc}
\hline
\hline
GRB & $t - t_0 \, (\mathrm{s})$ & $P \, (\%)$ & EVPA (deg) & Rank & Afterglow onset $t_\mathrm{peak} \, (\mathrm{s})$ & $\mathrm{T90}$ & $A_\mathrm{V}^\mathrm{Gal.}$ & $z$ \\
\hline
100805A & $140 - 320$ & $< 14$ & ... & 0.377 & $< 140.4$ & $16.7$ & $0.5$ & $\approx 1.3$ \\
101112A & $176 - 355$ & $6^{+3}_{-2}$  & $71\pm10$ & 0.978 & $299.0 \pm 6.0$ & $9.2$ & $0.4$ & $\lesssim 3.5$ \\
" & $715 - 893$ & $6^{+4}_{-3}$ & $76\pm15$ & 0.934 & &" &"&" \\
110205A & $240 - 840$ & $13^{+13}_{-9}$  & $126\pm26$ & 0.967 & $1027.0 \pm 8.0$ & $249$ & $0.04$ & $2.22$ \\
"& $3047 - 3645$ & $< 5$  & ... & 0.883 &" &  "&" &" \\
"& $3960 - 4140$ & $< 23$ & ... & 0.486 & "&" &" & "\\
110726A & $191 - 783$ & $< 14$ & ... & 0.331 & $< 191.2$ & $5.2$ & $0.21$ & $1.04 < z < 2.7$ \\
120119A & $194 - 793$ & $< 8$ & ... & 0.713 & $< 194.4$ & $68.0$ & $0.3$ & $1.728$ \\
120308A & 240 - 323 & $28 \pm 4$ & $34 \pm 4$ & $>0.99$ & $298.0 \pm 16.0$ & $61.3$ & $0.11$ & $2.22^{+0.25}_{-0.27}$ \\
" & 323 - 407 & $23 \pm 4$ & $44 \pm 6$ &  $>0.99$  & "& "&" &" \\
" & 407 - 491 & $17 ^{+4}_{-5}$ & $51 \pm 9$ &  $>0.99$  & "& "& "&" \\
" & 491 - 575 & $16 ^{+4}_{-7}$ & $40 \pm 10$ &  $>0.99$  & "&" &" &" \\
" & 575 - 827 & $16 ^{+4}_{-5}$ & $55 \pm 9$ &  $>0.99$ &" &" &"  &" \\
120311A & $181 - 779$ & $< 13$ & ... & 0.008 &$< 180.6$  & $3.5$ & $0.37$ & $\lesssim 3$ \\
120326A & $210 - 872$ & $< 14$ & ... & 0.139 &$< 216.0$  & $69.5$ & $0.14$ & $1.798$ \\
120327A & $1664 - 2263$ &  $< 4$ & ... & 0.505 &$< 1663.8$ & $63.5$ & $0.92$ & $2.81$ \\
"& $2605 - 2784$ &  $< 7$ & ... & 0.823 & "&" &" &"  \\
\hline
\end{tabular}
\end{center}

{\bf Notes.} Columns are: GRB identifier, interval of RINGO2 observations, degree of polarization, measured polarization sky angle (East of North), rank of the polarization measurement in permutation analysis, optical afterglow peak time, gamma-ray emission duration, Galactic extinction in V band, redshift.
\label{tab:sample_properties}
\end{table*}

We derived $q$ and $u$ values and their associated errors for all of the objects in all of the RINGO2
GRB frames following the procedure outlined in Section \ref{sec:pol-extract}.  These are presented in Figure  \ref{fig:qu-mag} as a function of apparent magnitude for all of the objects
in the sample (apart from GRB\,120308A where we have already presented such a plot in \cite{mundell-nature}).  The GRB measurements are indicated by red symbols.   In most cases the red symbols are within the general distribution of points, however in some cases (GRB\,10112A and 110205A) there is an apparent offset indicating a possible significant polarization detection.

To investigate the significance of these possible detections we needed to consider the  combination of information contained in both the $q$ and $u$ distributions.  To do this we used two independent and complementary methodologies. Our first method compared the calculated $P$ value (Equation 3) of the afterglow to all other sources in the field to test if the transient differs significantly from the population. Our second method used just the afterglow data itself and tested the null hypothesis that
the measured $P$ value was consistent with the scatter in its measured counts in the eight images
and therefore no detection could be claimed.

We note that the RINGO2 data on GRB\,110205A were originally presented in \cite{cucchiara} 
who gave an upper limit of $P<16$\% at the first epoch (240--840\,sec) and 
$P=3.6^{+2.6}_{-3.6}$ in the second epoch (3047--3645\,sec).  The enhanced analysis presented in the following sections is statistically consistent with these values but does formally give a marginal detection for the first epoch.   We also note that we originally presented results for GRB\,120308A in \cite{mundell-nature}.  In this case although we re-analysed the data along with the other bursts presented here, no changes to the results originally presented were found.  For that reason we do not plot the data for that burst in Figures \ref{fig:qu-mag}--\ref{fig:hist} as corresponding figures are already published there.  The results for this burst are however included in all of our tabular material and the discussion in Section \ref{sect:discussion}.

\subsection{Initial Analysis}
\label{sec:initial}

In Figure \ref{fig:stokes} we plot the Stokes $q$ versus $u$ parameter for all of the detected sources in each GRB observation.  For the GRB (red points) we also plot the error bars. The plots are generally characterized by a central ``blob'' of points near $q,u=0.0,0.0$ with the GRB often somewhat offset.  In most cases the error bars indicate the GRB is at least consistent with no polarization.  However the cases of GRB10112A and 110205A are not so clear, especially at their first epochs of observation.   In order to make an initial assessment of whether the offset of the GRB from the majority of other sources indicates a detection, we must take into account that every source will have different error bars (not plotted for clarity - see Figure \ref{fig:qu-mag} for the individual error bars on $q$ and $u$.)  

Since polarimetric error is a function of total counts for each individual source in the frame, we investigated the distribution of the measured polarization of every source in a particular frame as a function of magnitude.  For this analysis the polarimetric error on every source was calculated based on its individual $q$ and $u$ errors via a simple Monte Carlo error analysis.  As described in Section \ref{sec:pol-extract} the errors on $q$ and $u$ are expected to have a normal (Gaussian) distribution and can be calculated by standard photon counting statistics and error propagation theory \citep{clarke}.

For computational efficiency in this initial analysis the ratio $u/q$ was assumed to be fixed to its measured value for each source in the frame.   In other words we assumed the measured EVPA for each source in the frame was correct (although different for each source).  For each source a range of simulated $P$ values from 0.01\% to 70.00\% was then stepped through.  For each $P$ value, corresponding $q$ and $u$ values were calculated (based on the EVPA).  The Python numpy.randn Gaussian weighted random number generator was then used to generate
two separate 1000 value distributions centred on the calculated $q$ and $u$ values with standard deviations equal to the error estimated on each quantity calculated using \cite{clarke}.
These $q$ and $u$ distributions were then combined using Equation (3) to calculate a simulated $P$ distribution which was examined to see if the observed $P$ lay within its 1$\sigma$ limits and therefore the simulated $P$ was ``valid''.  The maximum and minimum valid $P$ values after stepping through the whole range of simulated $P$ therefore gave the error bars for that particular source.  Tests showed this procedure gives identical results to the graphical method presented in \cite{simmons}.  

The results of this analysis are presented in Figure \ref{fig:trumpet}.  As expected the figure shows
an increase in polarimetric error for fainter sources.  However it can also be seen that the
apparent measured $P$ value (calculated via Equation 3) also increases for fainter sources.  This is not real, and is an example of polarimteric bias.  As signal-to-noise ratio decreases, the noise is converted to signal via the non-linear nature of Equation 3.

Figure \ref{fig:trumpet} shows that in general the GRBs are located within the expected noise at their measured magnitude level and therefore no claim of polarization detection can be made.  However, both epochs of GRB\,101112A and the first epoch of 110205A show the GRB located offset from the general cloud of points, indicating a possible polarization detection.

\subsection{Permutation Analysis of Detection Probabilities}

In order to investigate the detection probabilities more thoroughly, we therefore used our second method.  This is a ``permutation analysis'' of the set of 8 measured counts for each GRB prior to  their
conversion to Stokes parameters.  To do this we first  had to remove the imprinted signal of instrumental polarization from  the 
measured counts. This signal can be characterized by a response array of 8 values.  It is calculated for each of the 8 rotor positions by averaging the normalized counts in all of our observations of zero polarized standards to create a 8 value response array.  Dividing this 8 value response array into the measured 8 count values for the GRB
then removes the effect of instrumental  polarization (see \cite{ringo} for more details of this alternative approach to instrumental polarization correction to that done in the $qu$ plane.)

Following this, we constructed all (8-1)! permutations of the ordering of the corrected count values to generate 5040 different sets of 8 flux values.  These sets have similar noise characteristics to the original data, being constructed directly from it.    We then measured the polarization degree from each of these sets, and computed the ranking of the measured polarization of the GRB within all of the sets constructed from that GRBs reordered data.  
If the measured polarization was simply a result of noise superimposed on a zero polarized object, we would expect the measured polarization to lie randomly within the distribution of polarization values.  This is therefore a test of the hypothesis that the polarization signal is non-zero.  

The results of our analysis are presented in Figure \ref{fig:hist} and Table \ref{tab:sample_properties}.  The ranks expressed as a fraction of the total number of permutations are equivalent to a probability that the measured GRB polarization is not the result of random noise. All epochs of measurement in GRB120308A, the first and second epoch of GRB\,101112A and the first epoch of GRB\,110205A have a probability $p(=1-rank)$<0.1 of being consistent with zero percent polarization. No other bursts have detections that are not consistent with zero polarization at our confidence limit.  This result is entirely consistent with the results from our first method analysis, and gives us confidence in our approach.  We note that  (putting aside  GRB120308A for  which the non-zero polarization 
confidence is very  high) we have  3 out of 12 measurements with $p<0.1$.  A binomial
analysis shows the probability of zero polarization for all three measurements is 11\% making the conservative assumption of $p=0.1$ in all cases.

\subsection{Final Polarization Values}

To determine the final polarization values and error bars for our polarization detections of the GRB optical counterparts, and the upper limits in the case where no positive polarization detection could be made, we carried out a more sophisticated version of the Monte Carlo analysis from our first method.  In this case we relaxed the constraint requiring that the ratio $u/q$ is fixed.  This assumption is particularly poor when the $u$ and $q$ error bars approach the origin and was imposed in the ``all objects in all frames'' analysis of Section \ref{sec:initial} due to computational constraints.  In this final analysis we therefore explored simulated ranges of polarization from 0.01\% to 70.00\% and EVPA from 0.0$^{\circ}$ to 179.9$^{\circ}$.  The mean of the distribution of "valid" polarization and EVPA values were then taken as the final measured value (as opposed to simply applying Equations 3 and 4).  This procedure corrects for polarization bias \citep{jermak-mnras,simmons} although the corrections are in any case small ($<1$\%) compared to the measured values and their associated errors.  The 10\% and 90\% limits of the distribution were used to define the error bars.  For any particular burst, if the lower error bar reached zero, the upper error bar was interpreted as an upper limit.  The results of this procedure are presented in Table \ref{tab:sample_properties}.

\section{Discussion}
\label{sect:discussion}

\renewcommand{\arraystretch}{1.25}
\begin{table}[!h]\footnotesize
\begin{center}
\caption{Light curve fitting results.}
\label{tab:model_results}
\begin{tabular}{llll}
\hline
\hline
GRB & Model$^\dagger$ & Fit parameters & $\chi ^2$ (d.o.f.) \\
\hline
100805A & PL & $\alpha_\mathrm{decay}^\mathrm{PL} = 0.86 \pm 0.04$ & $30.8$ ($24$) \\
101112A & B & $\alpha_\mathrm{rise} = -4.24 \pm 2.95$ & $19.4$ ($28$) \\
& & $\alpha _\mathrm{decay} = 1.10 \pm 0.05$ & \\
& & $t _\mathrm{peak} = 299 \pm 6 \,\mathrm{s}$ & \\
& & $n = 0.86 \pm 0.72$ & \\
110205A & B & $\alpha_\mathrm{rise} = -4.63 \pm 0.29$ & $220.5$ ($84$) \\
& & $\alpha _\mathrm{decay} = 1.52 \pm 0.02$ & \\
& & $t _\mathrm{peak} = 1027 \pm 8 \,\mathrm{s}$ & \\
& & $n = 2.18 \pm 0.45$ & \\
110726A & PL + B & $\alpha_\mathrm{decay}^\mathrm{PL} = 1.03 \pm 0.05$ & $32.1$ ($30$) \\
& & $\alpha _\mathrm{rise} = -7.87 \pm 21.21$ & \\
& & $\alpha _\mathrm{decay} = 1.13 \pm 0.33$ & \\
& & $t _\mathrm{peak} = 3256 \pm 185 \,\mathrm{s}$ & \\
& & $n = 0.40 \pm 1.23$ & \\
120119A & PL + B & $\alpha_\mathrm{decay}^\mathrm{PL} = 0.65 \pm 0.06$ & $105.8$ ($74$) \\
& & $\alpha _\mathrm{rise} = -1.06 \pm 0.41$ & \\
& & $\alpha _\mathrm{decay} = 1.68 \pm 0.19$ & \\
& & $t _\mathrm{peak} = 822 \pm 22 \,\mathrm{s}$ & \\
& & $n = 1.05 \pm 0.48$ & \\
120308A & B + B$^\ast$ & $\alpha_\mathrm{rise}^1 = -5$ & $10.7$ ($17$) \\
& & $\alpha _\mathrm{decay}^1 = 2.4 \pm 0.6$ & \\
& & $t _\mathrm{peak}^1 = 298 \pm 16 \,\mathrm{s}$ & \\
& & $n^1 = 1$ & \\
& & $\alpha _\mathrm{rise}^2 = -0.5$ & \\
& & $\alpha _\mathrm{decay}^2 = 1.4 \pm 0.1$ & \\
& & $t _\mathrm{peak}^2 = 730^{+190} _{-150} \,\mathrm{s}$ & \\
& & $n^2 = 1$ & \\
120311A & PL & $\alpha_\mathrm{decay}^\mathrm{PL} = 1.03 \pm 0.06$ & $12.7$ ($13$) \\
120326A & PL & $\alpha_\mathrm{decay}^\mathrm{PL} = 0.42 \pm 0.04$ & $12.9$ ($12$) \\
120327A & PL & $\alpha_\mathrm{decay}^\mathrm{PL} = 1.22 \pm 0.02$ & $25.2$ ($50$) \\
\hline
\end{tabular}
\end{center}

{\bf Notes.} $^\dagger$PL is a simple power-law model ($F \propto t^{-\alpha}$, while B is a Beuermann model (smoothly joint broken power-law model, see \citealt{beuermann1999}). $^\ast$Results from Mundell et al. (2013).
\end{table}

\begin{figure}[!h]
\begin{center}
\includegraphics[width=1\linewidth]{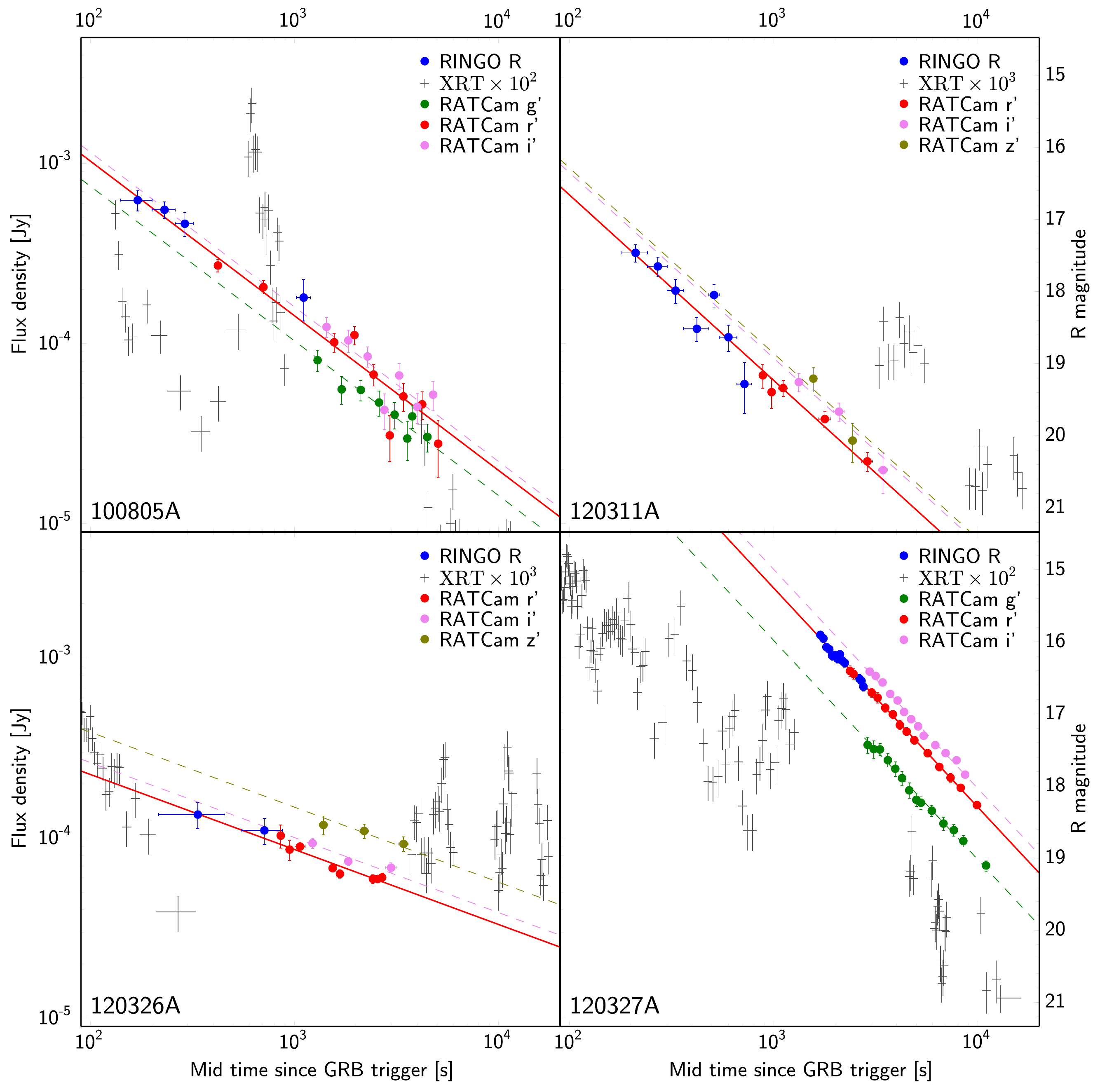}
\caption{\label{fig:lc1} Light curves for GRB\,100805A, GRB\,120311A, GRB\,120326A and GRB\,120327A, which show single power-law decay morphology.}
\end{center}
\end{figure}

\begin{figure}[!h]
\begin{center}
\includegraphics[width=1\linewidth]{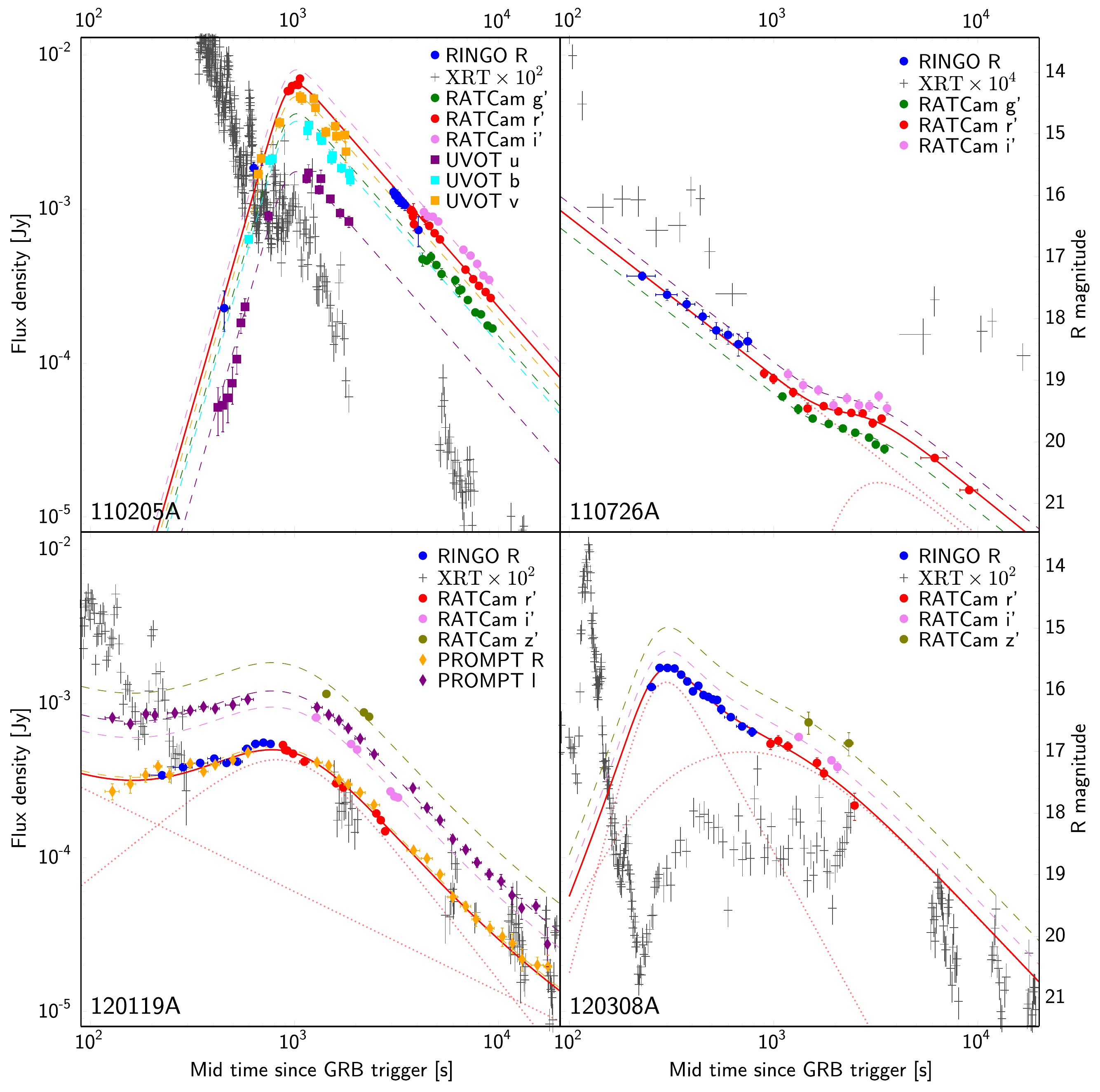} 
\caption{\label{fig:lc2} GRB\,110205A, GRB\,110726A, GRB\,120119A and GRB\,120308A for which the light curves show definite structure. The Beuermann and power law components defined in Table~\ref{tab:model_results} are plotted individually as dotted lines and the final r-band model fit (the summation of the multiple components) plotted as a solid line. To more easily compare by eye the light curve shape between filters, the model is plotted multiple times offset to align with the non-$r'$ band filters and plotted as a dashed line. The steep rise for GRB\,110205A and GRB\,120308A indicates the presence of the reverse-shock component in the afterglow.}
\end{center}
\end{figure}

\begin{figure}[!h]
\begin{center}
\includegraphics[width=1\linewidth]{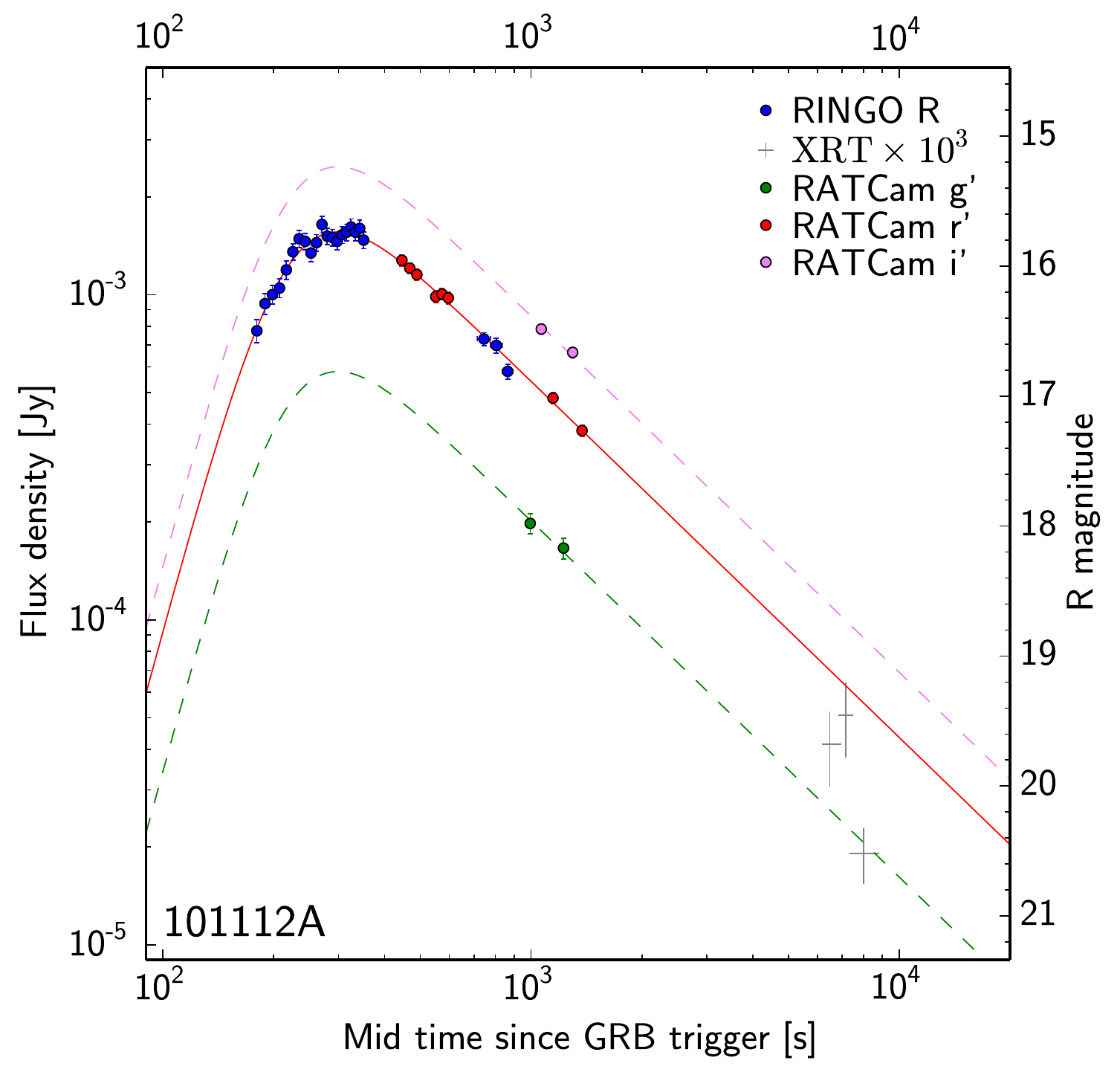} 
\caption{\label{fig:lc101112} GRB\,101112A light curve. Steep rise indicates the presence of the reverse-shock component, while the shallow decay indicates that reverse-shock and forward-shock components peak at similar times.}
\end{center}
\end{figure}

We fitted the optical light curves of our nine afterglows with a simple power-law (PL) 
or/and a smoothly jointed broken power-law function (B) \citep{beuermann1999}.   We followed the
fitting procedure outlined in \citep{kopac2013} where for
each GRB we start by fitting a simple power-law, and then if the fit is not satisfactory, we add additional components: firstly a broken power-law (B), then B + single PL, and then finally 2 Beuermann functions.  We always fit the complete optical dataset simultaneously
(i.e all the filters at the same time but assuming no color evolution, i.e. only a normalization change for each filter). In the case of the combined functions a simple linear addition of the two components is made, with their relative contributions normalized via the PL fit parameters.

The fitting results (e.g., decay and rising indexes and peak 
times) are summarized in Table~\ref{tab:model_results}. The light curves of four events: GRB 100805A, 
GRB 120311A, GRB 120326A, and GRB 120327A are well modeled by a simple PL function
as shown in Figure \ref{fig:lc1}. Although GRB 120326A indicates a very shallow decay with 
$\alpha_{decay} \sim 0.42$ (possibly due to refreshed shocks), the others are 
consistent with the standard forward shock emission $\alpha_{decay} \sim 1$ \citep{Sari1998}. For these events, the duration $T_{90}$ of the prompt 
gamma-rays are $3.5-70$ s, and the optical observations started well after 
the end of the prompt gamma-ray emission phases. The observations were 
not prompt enough to detect the onset of afterglow and the optical emission is 
dominated by the forward shock emission in these observations. 
Since the forward shock region is expected to contain only highly tangled 
magnetic fields generated around the shock (\cite{med99}; 
however, see also \citet{uehara2012}), the non-detection of polarization 
is consistent with the forward shock model. 
Also plotted in Figure \ref{fig:lc1} are the X-ray light curves (black crosses).
They indicate significant, multiple flares in the early phase. These X-ray 
flares have been reported in many events, and the rapid variability $\Delta t/t$
indicates that these originate from internal dissipation processes, rather than 
forward shock e.g. \citep{zhang2006,nousek2006}.

The other five events show more complicated behavior in the early optical afterglow as shown in Figures \ref{fig:lc2} and \ref{fig:lc101112}. These light curves indicate a peak or/and re-brightening at later times. The three events for which we have detected polarization signals are all in this group:
\begin{itemize}
\item GRB 101112A: we detected $\sim 6\%$ 
polarization degree around the peak and in the decay phase $\alpha_{decay} \sim 1.1$.
If the peak at $t_{peak} \sim 299$ s is the onset of afterglow, considering 
$t_{peak} \gg T_{90}\sim 9.2 ~\mbox{s}$, this is a thin shell case
\citep{saripiran95,k99}. The expected rising $t^{3}$ 
of the (slow-cooling) forward shock emission is slower than the observed 
rising $\sim t^{4.2}$, and it implies that the reverse shock emission contributed 
around the peak \citep{kobayashi2000}. Although the fast-cooling forward shock emission can rise as rapidly as $\sim t^{3.7}$, the expected decay $t^{-1/4}$ after the peak due to 
the passage of the cooling frequency is very shallow, and it is not consistent with 
our observations. 
\item GRB 110205A: the peak at $t_{peak}\sim 1027$ s ($\gg T_{90}=249$ s) 
is considered to be the onset-of the afterglow. The rapid rise $t^{4.6}$ and decay $t^{-1.5}$
implies the contribution to the peak from a reverse shock in the thin shell regime. 
A polarization degree of $13\%$ was detected in the rising phase.
\item GRB 120308A: 
we detected polarization degrees as high as $28\%$ for this event. The high polarization 
was detected around the peak at $t_{peak}=298$ s ($\gg T_{90}=61.3$ s), and the very rapid 
rise $t^5$ and decay $t^{-2.4}$ are a clear signature of the reverse shock. 
In \citet{mundell-nature} we demonstrated that this light curve is best described
by the combination of the two components, one from a reverse and the other from a forward shock.
\end{itemize}
We also note we detected polarization in multiple epochs for GRB 101112A and GRB120308A, with constant EVPA within the error limits in both cases. 

Polarization signals were not detected from the remaining two events which show 
peak or/and re-brightening in their afterglow light curves:
\begin{itemize}
\item GRB 110726A: the light curve initially decays with $\alpha_{decay}=1.03$, and it shows a 
re-brightening around $t=3200$ s. The polarization limits were obtained during the 
initial power-law decay phase. The decay index is consistent with the forward shock emission. 
Except for the re-brightening which is possibly due to energy injection \citep{zhang2006,nousek2006}, this event looks similar to the power-law events 
shown in Figure 9. 
\item GRB 120119A: a broad peak is noticeable in the light curve. The rise 
is very slow, the I band light curve is almost flat at the beginning. The polarization 
limit was obtained during the slow rising phase. This broad peak can be 
reasonably explained by forward shock models with energy injection or density enhancement 
in the ambient medium \citep{zhang2006,nousek2006}. 
\end{itemize}

\begin{figure}[!h]
\begin{center}
\includegraphics[width=0.95\linewidth]{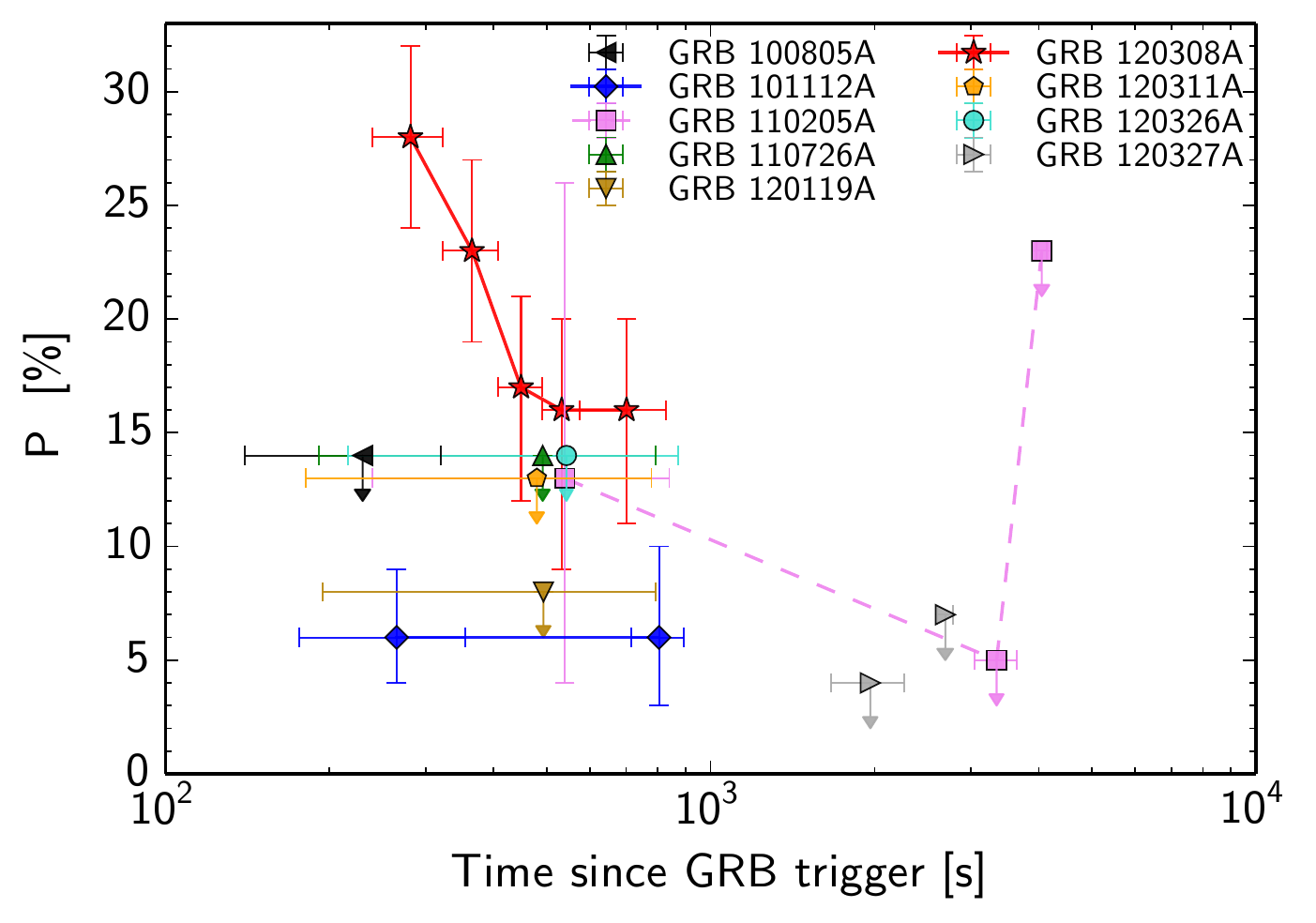} 
\caption{\label{fig:polvstime}Polarization degree as a function of time in after the burst for all $9$ GRBs from the RINGO2 sample. The temporal error bars show the duration of the exposure. }
\end{center}
\end{figure}

Figure \ref{fig:polvstime} shows the polarization measurements (detection or upper-limit) 
of all nine events as a function of the observing time since the GRB trigger. 
We note that all polarization detection cases (GRB101112A, GRB110205A and 
GRB120308A) were achieved at relatively 
early times $t< 10^3$ s. This reinforces the point that prompt measurements are essential to characterize the polarimetric properties of GRB afterglow; the 
polarization degree decays very rapidly as the tight upper-limits at late times show. 

\begin{figure}[!h]
\begin{center}
\includegraphics[width=0.95\linewidth]{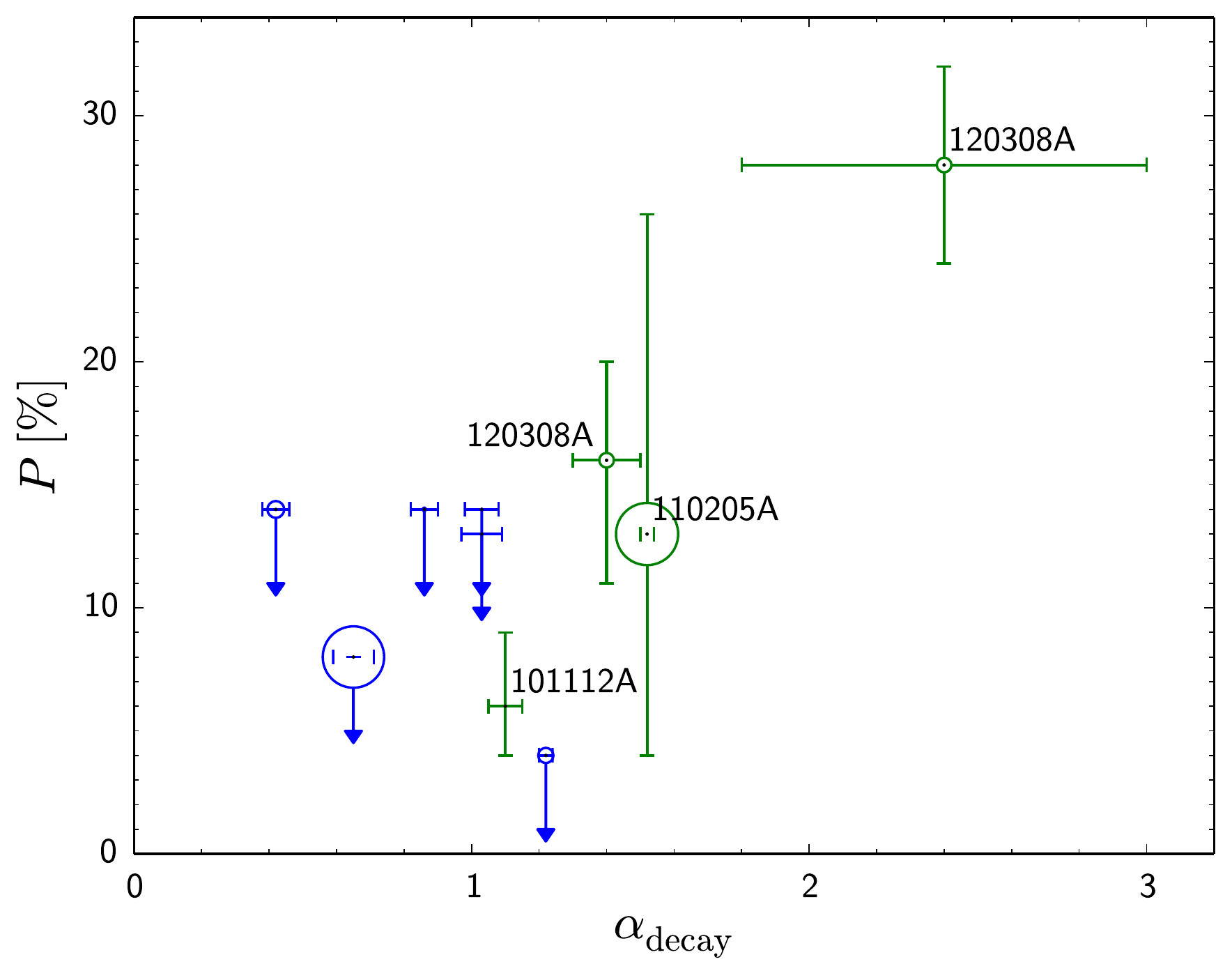} 
\caption{\label{fig:alpha_vs_P}.  Power law decay index ($\alpha$) versus degree of Polarization. The 
green points on the plot are measurements, while blue points are 
upper limits. The relative size of the point is the T90 value (which  
shows no correlation with $\alpha$ or and $P$).  For GRB120308A we plot
two epochs: ($240-323$ sec, $P=28$\%) and ($575-827$ sec, $P=16$\%).
}
\end{center}
\end{figure}

All polarized cases suggest the reverse shock emission at early times. 
Since no new electrons are shocked after the reverse shock has crossed GRB ejecta, 
the reverse shock emission is short lived and it decays faster than the emission 
from the forward shock which continuously shocks electrons in ambient medium. 
Therefore, a rapid decay, typically $t^{-2}$, is also a signature of the reverse shock
emission \citep{SariPiran1999,kobayashi2000,zhang2003,japelj2014}. 
We therefore tested the correlation between the observed decay index and polarization degree. 
Figure \ref{fig:alpha_vs_P} shows that the polarized cases (the green crosses) do indeed have larger 
decay indexes.  The light curve of GRB120308A shows a double peak structure with
reverse and forward shock peaks at different times.  The polarization degree is 
much higher during the clearly separated reverse shock peak. However, 
for GRB 110205A, the polarization $P=13\%$ is detected only in the rising phase, 
and we have a tight upper limits of $P < 5\%$ in the decay phase ($\alpha_{decay}=1.52$). 

\citet{zheng2012} showed that the full optical and x-ray afterglow of GRB 110205A
could be interpreted within the standard reverse shock + forward shock model, and 
they proposed two scenarios. Scenario I 
invokes both the forward shock and reverse shock to peak at $\sim 10^3$ s, while 
Scenario II invokes the reverse shock only to the peak at $\sim 10^3$ s, with the 
forward shock peak later when the typical frequency crosses the optical band. 
According to their modeling (see Figure 5 in their paper), the reverse shock 
contribution becomes negligible by our polarization observations around 3000-3600 s. 
Our limit $P<5\%$ is consistent with the dominance of the forward shock emission 
in the optical band. In Scenario II, the optical band is still dominated by 
the reverse shock emission in the observation period. 
Because of the relativistic beaming effect, we can see 
only a small portion of the GRB ejecta just around the line of sight with angular scale 
of $1/\Gamma_0 \sim 4\times 10^{-3}$ where $\Gamma_0 \sim 250$ is 
the initial Lorentz factor of the ejecta \citep{zheng2012}. After the reverse 
shock crossing, the ejecta rapidly decelerates as $\Gamma \propto R^{-g} \sim 
R^{-2}$ in terms of the ejecta radius (Kobayashi \& Sari 2000). However, it is 
not so rapid in terms of the time $\Gamma \propto t^{-g/(1+2g)} \propto t^{-0.4}$. 
By $t \sim 3000$ s, the angular size of the visible region grows only by a factor 
of $\sim 3^{0.4}\sim 1.6$, compared to the size at the peak time $t \sim 10^3$ s. 
Although a larger visible region at a later time potentially reduces the polarization 
degree if the magnetic fields have an irregularity in the angular scale of $1/\Gamma_0$ 
or a slightly larger scale,  this small change in the size does not explain the drastic change 
from $P=13\%$ to $P<5\%$. Our polarization measurements therefore disfavor Scenario II.

\par

\section{Conclusion}
\label{sect:conclusion}

We have presented the complete RINGO2 catalog of GRB afterglow observations. 
We carried out 19 prompt RINGO2 observations between 2010 and 2012.
9 out of the 19 events were bright enough to perform polarimetric analysis, the polarization 
degrees (or limits) and EVPA were measured. We detected polarization signals in 
their early optical afterglow for three events: GRB 101112A, GRB110205A and GRB120308A.
Using RINGO2 and RATCam data, we constructed the 
light curves of the bright events to evaluate the decay indexes of the afterglow.
The combination of our photometric and polarimetric data have shown that there is a correlation between decay index and polarization degree, i.e. polarized events decay faster.  It clearly indicates that the events for which polarization were detected have a reverse shock emission component in the early afterglow. 

The internal energy produced by shocks is believed to be radiated via synchrotron 
emission. The presence of strong magnetic fields is crucial in the standard 
synchrotron shock model. Although magnetic fields are usually assumed to be 
generated locally by instabilities in shocks, with the resultant tangled fields,
the polarization signals are canceled out. The polarized 
reverse shock emission indicates that there are large scale magnetic fields in 
the original GRB ejecta which are likely to be generated at the GRB central engine. 
We have detected polarization signals in multiple epochs for two events:  
GRB 101112A and GRB120308A. In the former case, the polarization degree is 
constant around the onset of afterglow and in the decay phase. 
The latter shows the gradual decay of the polarization signals.
EVPA remains constant within the error limits in both cases. 
In magnetic GRB jet models that assume the amplification of magnetic fields by 
the rotation of the central black hole and the accretion disk, the outflow is 
expected to be thread with globally ordered magnetic fields which is likely to 
be dominated by a toroidal component, 
because the radial field decays faster than the tangential one. Although the
toroidal fields can be distorted by internal dissipation processes 
preceding the onset of afterglow e.g. \cite{ZhangYan2011}, the visible region with 
angular scale $\sim 1/\Gamma$ might have a rather uniform magnetic field and the polarization (electric) vector is expected to point toward the jet axis. The constant EVPA results are consistent with this model. 

As illustrated especially in the case of GRB 110205A and GRB 120308A, polarimetry 
allowed us to carry out the detailed modeling of early afterglow. Polarization measurements can distinguish the forward shock and reverse shock emission components. Since the reverse shock emission is short lived, prompt polarization measurements at less than $t \sim 10^3$ s
are essential to fully characterize the early afterglow and constrain the GRB 
central engine \citep{kopac15}. 

\acknowledgments{
The Liverpool Telescope is operated on the island of La Palma by Liverpool John Moores University in the Spanish Observatorio del Roque de los Muchachos of the Instituto de Astrofisica de Canarias with financial support from the UK Science and Technology Facilities Council (STFC). CGM acknowledges support from the Royal Society, the Wolfson Foundation and STFC.
}

\newpage

\appendix
\label{sec:appendix}
\renewcommand\arraystretch{1.3}
\begin{longtable}{cccccc}
\caption{RINGO2 sample complete photometry. Magnitudes are corrected for the Galactic extinction.} \label{tab:photometry} \\
\hline
\hline 
GRB & $t_\mathrm{start}$ & Exp & Filter & Magnitude & $F_\nu^\mathrm{OPT}$ \\
 & [min] & [s] & & & [mJy] \\
\hline
\endfirsthead

\multicolumn{6}{c}%
{\bfseries \tablename\ \thetable\ -- continued from previous page}\\
\hline 
\hline
GRB & $t_\mathrm{start}$ & Exp & Filter & Magnitude & $F_\nu^\mathrm{OPT}$ \\ 
 & [min] & [s] & & & [mJy] \\
\hline 
\endhead

\\
\hline
\multicolumn{6}{r}{{Continued on next page}} \\
\hline
\endfoot

\\ \hline \hline
\endlastfoot

100805A & $2.34$ & $60.6$ & RINGO r' & $17.29 \pm 0.13$ & $0.653 \pm 0.078$ \\
 & $3.35$ & $59.9$ & RINGO r' & $17.42 \pm 0.11$ & $0.579 \pm 0.061$ \\
 & $4.35$ & $58.7$ & RINGO r' & $17.62 \pm 0.16$ & $0.484 \pm 0.069$ \\
 & $16.99$ & $178.9$ & RINGO r' & $18.66 \pm 0.28$ & $0.190 \pm 0.048$ \\
 & $6.78$ & $30.0$ & RATCam r' & $18.24 \pm 0.09$ & $0.271 \pm 0.021$ \\
 & $11.49$ & $30.0$ & RATCam r' & $18.54 \pm 0.09$ & $0.206 \pm 0.017$ \\
 & $25.82$ & $30.0$ & RATCam r' & $19.31 \pm 0.13$ & $0.102 \pm 0.012$ \\
 & $32.53$ & $30.0$ & RATCam r' & $19.21 \pm 0.13$ & $0.111 \pm 0.013$ \\
 & $40.18$ & $60.0$ & RATCam r' & $19.76 \pm 0.15$ & $0.067 \pm 0.009$ \\
 & $48.33$ & $60.0$ & RATCam r' & $20.64 \pm 0.32$ & $0.031 \pm 0.009$ \\
 & $56.43$ & $60.0$ & RATCam r' & $20.07 \pm 0.19$ & $0.051 \pm 0.009$ \\
 & $69.38$ & $120.0$ & RATCam r' & $20.18 \pm 0.19$ & $0.046 \pm 0.008$ \\
 & $82.70$ & $180.0$ & RATCam r' & $20.78 \pm 0.40$ & $0.028 \pm 0.010$ \\
 & $21.38$ & $30.0$ & RATCam g' & $19.75 \pm 0.15$ & $0.081 \pm 0.011$ \\
 & $28.12$ & $30.0$ & RATCam g' & $20.16 \pm 0.19$ & $0.056 \pm 0.010$ \\
 & $34.73$ & $60.0$ & RATCam g' & $20.16 \pm 0.14$ & $0.055 \pm 0.007$ \\
 & $42.76$ & $60.0$ & RATCam g' & $20.34 \pm 0.17$ & $0.047 \pm 0.008$ \\
 & $51.05$ & $60.0$ & RATCam g' & $20.51 \pm 0.18$ & $0.040 \pm 0.007$ \\
 & $59.00$ & $60.0$ & RATCam g' & $20.86 \pm 0.28$ & $0.030 \pm 0.007$ \\
 & $61.92$ & $120.0$ & RATCam g' & $20.53 \pm 0.16$ & $0.039 \pm 0.006$ \\
 & $73.11$ & $180.0$ & RATCam g' & $20.82 \pm 0.19$ & $0.030 \pm 0.005$ \\
 & $23.68$ & $30.0$ & RATCam i' & $18.99 \pm 0.14$ & $0.124 \pm 0.016$ \\
 & $30.33$ & $30.0$ & RATCam i' & $19.18 \pm 0.16$ & $0.104 \pm 0.015$ \\
 & $37.53$ & $60.0$ & RATCam i' & $19.40 \pm 0.14$ & $0.085 \pm 0.011$ \\
 & $45.40$ & $60.0$ & RATCam i' & $20.16 \pm 0.25$ & $0.043 \pm 0.010$ \\
 & $53.78$ & $60.0$ & RATCam i' & $19.67 \pm 0.18$ & $0.067 \pm 0.011$ \\
 & $65.57$ & $120.0$ & RATCam i' & $20.11 \pm 0.21$ & $0.045 \pm 0.009$ \\
 & $78.10$ & $180.0$ & RATCam i' & $19.94 \pm 0.20$ & $0.052 \pm 0.010$ \\
101112A & $2.94$ & $7.2$ & RINGO r' & $16.99 \pm 0.09$ & $0.801 \pm 0.066$ \\
 & $3.09$ & $8.1$ & RINGO r' & $16.77 \pm 0.08$ & $0.980 \pm 0.073$ \\
 & $3.24$ & $8.1$ & RINGO r' & $16.71 \pm 0.07$ & $1.035 \pm 0.069$ \\
 & $3.39$ & $8.1$ & RINGO r' & $16.66 \pm 0.11$ & $1.087 \pm 0.107$ \\
 & $3.54$ & $8.1$ & RINGO r' & $16.52 \pm 0.06$ & $1.232 \pm 0.073$ \\
 & $3.69$ & $8.1$ & RINGO r' & $16.38 \pm 0.06$ & $1.402 \pm 0.083$ \\
 & $3.84$ & $8.1$ & RINGO r' & $16.28 \pm 0.11$ & $1.542 \pm 0.152$ \\
 & $3.99$ & $8.1$ & RINGO r' & $16.30 \pm 0.06$ & $1.509 \pm 0.079$ \\
 & $4.14$ & $8.1$ & RINGO r' & $16.39 \pm 0.06$ & $1.389 \pm 0.082$ \\
 & $4.29$ & $8.1$ & RINGO r' & $16.30 \pm 0.12$ & $1.515 \pm 0.163$ \\
 & $4.44$ & $8.1$ & RINGO r' & $16.16 \pm 0.11$ & $1.723 \pm 0.170$ \\
 & $4.59$ & $8.1$ & RINGO r' & $16.25 \pm 0.08$ & $1.582 \pm 0.117$ \\
 & $4.74$ & $8.1$ & RINGO r' & $16.27 \pm 0.06$ & $1.552 \pm 0.091$ \\
 & $4.89$ & $8.1$ & RINGO r' & $16.30 \pm 0.08$ & $1.511 \pm 0.112$ \\
 & $5.04$ & $8.1$ & RINGO r' & $16.25 \pm 0.06$ & $1.580 \pm 0.082$ \\
 & $5.19$ & $8.1$ & RINGO r' & $16.23 \pm 0.08$ & $1.611 \pm 0.119$ \\
 & $5.34$ & $8.1$ & RINGO r' & $16.19 \pm 0.09$ & $1.673 \pm 0.138$ \\
 & $5.49$ & $8.1$ & RINGO r' & $16.23 \pm 0.10$ & $1.614 \pm 0.146$ \\
 & $5.64$ & $8.1$ & RINGO r' & $16.20 \pm 0.09$ & $1.658 \pm 0.136$ \\
 & $5.79$ & $7.2$ & RINGO r' & $16.29 \pm 0.06$ & $1.523 \pm 0.090$ \\
 & $11.91$ & $60.4$ & RINGO r' & $17.05 \pm 0.05$ & $0.756 \pm 0.035$ \\
 & $12.91$ & $60.4$ & RINGO r' & $17.09 \pm 0.08$ & $0.730 \pm 0.054$ \\
 & $13.92$ & $58.1$ & RINGO r' & $17.30 \pm 0.06$ & $0.601 \pm 0.031$ \\
 & $7.35$ & $10.0$ & RATCam r' & $16.48 \pm 0.04$ & $1.278 \pm 0.053$ \\
 & $7.72$ & $10.0$ & RATCam r' & $16.54 \pm 0.04$ & $1.209 \pm 0.046$ \\
 & $8.08$ & $10.0$ & RATCam r' & $16.59 \pm 0.04$ & $1.155 \pm 0.048$ \\
 & $9.10$ & $10.0$ & RATCam r' & $16.76 \pm 0.04$ & $0.987 \pm 0.041$ \\
 & $9.47$ & $10.0$ & RATCam r' & $16.74 \pm 0.04$ & $1.006 \pm 0.041$ \\
 & $9.85$ & $10.0$ & RATCam r' & $16.77 \pm 0.04$ & $0.978 \pm 0.040$ \\
 & $18.87$ & $30.0$ & RATCam r' & $17.54 \pm 0.04$ & $0.481 \pm 0.020$ \\
 & $22.68$ & $30.0$ & RATCam r' & $17.79 \pm 0.04$ & $0.382 \pm 0.016$ \\
 & $16.33$ & $30.0$ & RATCam g' & $18.66 \pm 0.08$ & $0.198 \pm 0.014$ \\
 & $20.17$ & $30.0$ & RATCam g' & $18.85 \pm 0.08$ & $0.166 \pm 0.012$ \\
 & $17.52$ & $30.0$ & RATCam i' & $16.92 \pm 0.04$ & $0.785 \pm 0.030$ \\
 & $21.38$ & $30.0$ & RATCam i' & $17.10 \pm 0.04$ & $0.665 \pm 0.027$ \\
110205A & $7.04$ & $60.0$ & RINGO r' & $18.11 \pm 0.32$ & $0.223 \pm 0.064$ \\
 & $10.04$ & $60.0$ & RINGO r' & $15.78 \pm 0.09$ & $1.832 \pm 0.159$ \\
 & $11.04$ & $60.0$ & RINGO r' & $15.63 \pm 0.11$ & $2.105 \pm 0.205$ \\
 & $50.43$ & $60.0$ & RINGO r' & $16.20 \pm 0.09$ & $1.244 \pm 0.108$ \\
 & $51.43$ & $60.0$ & RINGO r' & $16.27 \pm 0.09$ & $1.166 \pm 0.096$ \\
 & $52.43$ & $60.0$ & RINGO r' & $16.25 \pm 0.09$ & $1.188 \pm 0.098$ \\
 & $53.43$ & $60.0$ & RINGO r' & $16.35 \pm 0.09$ & $1.084 \pm 0.094$ \\
 & $54.43$ & $60.0$ & RINGO r' & $16.34 \pm 0.09$ & $1.093 \pm 0.090$ \\
 & $55.43$ & $60.0$ & RINGO r' & $16.38 \pm 0.09$ & $1.054 \pm 0.091$ \\
 & $56.43$ & $60.0$ & RINGO r' & $16.38 \pm 0.09$ & $1.054 \pm 0.087$ \\
 & $57.43$ & $60.0$ & RINGO r' & $16.41 \pm 0.10$ & $1.026 \pm 0.094$ \\
 & $65.97$ & $178.8$ & RINGO r' & $16.44 \pm 0.51$ & $1.104 \pm 0.480$ \\
 & $15.37$ & $10.0$ & RATCam r' & $14.52 \pm 0.01$ & $5.835 \pm 0.059$ \\
 & $15.75$ & $10.0$ & RATCam r' & $14.51 \pm 0.01$ & $5.889 \pm 0.060$ \\
 & $16.13$ & $10.0$ & RATCam r' & $14.44 \pm 0.01$ & $6.293 \pm 0.075$ \\
 & $16.92$ & $10.0$ & RATCam r' & $14.42 \pm 0.01$ & $6.416 \pm 0.077$ \\
 & $17.28$ & $10.0$ & RATCam r' & $14.42 \pm 0.01$ & $6.392 \pm 0.071$ \\
 & $17.65$ & $10.0$ & RATCam r' & $14.32 \pm 0.02$ & $7.035 \pm 0.097$ \\
 & $61.80$ & $10.0$ & RATCam r' & $16.46 \pm 0.05$ & $0.978 \pm 0.041$ \\
 & $62.17$ & $10.0$ & RATCam r' & $16.45 \pm 0.05$ & $0.985 \pm 0.044$ \\
 & $62.53$ & $10.0$ & RATCam r' & $16.46 \pm 0.06$ & $0.976 \pm 0.053$ \\
 & $63.30$ & $10.0$ & RATCam r' & $16.56 \pm 0.08$ & $0.894 \pm 0.069$ \\
 & $63.68$ & $10.0$ & RATCam r' & $16.52 \pm 0.17$ & $0.938 \pm 0.147$ \\
 & $64.05$ & $10.0$ & RATCam r' & $16.68 \pm 0.10$ & $0.798 \pm 0.074$ \\
 & $72.58$ & $30.0$ & RATCam r' & $16.58 \pm 0.10$ & $0.875 \pm 0.082$ \\
 & $75.98$ & $30.0$ & RATCam r' & $16.71 \pm 0.01$ & $0.778 \pm 0.009$ \\
 & $80.45$ & $60.0$ & RATCam r' & $16.83 \pm 0.01$ & $0.697 \pm 0.004$ \\
 & $85.40$ & $60.0$ & RATCam r' & $16.93 \pm 0.03$ & $0.636 \pm 0.016$ \\
 & $113.63$ & $120.0$ & RATCam r' & $17.42 \pm 0.05$ & $0.404 \pm 0.019$ \\
 & $123.55$ & $180.0$ & RATCam r' & $17.57 \pm 0.01$ & $0.351 \pm 0.004$ \\
 & $132.45$ & $120.0$ & RATCam r' & $17.68 \pm 0.01$ & $0.317 \pm 0.003$ \\
 & $142.40$ & $180.0$ & RATCam r' & $17.78 \pm 0.01$ & $0.289 \pm 0.004$ \\
 & $151.35$ & $120.0$ & RATCam r' & $17.88 \pm 0.01$ & $0.265 \pm 0.002$ \\
 & $70.35$ & $30.0$ & RATCam g' & $17.27 \pm 0.11$ & $0.473 \pm 0.047$ \\
 & $73.72$ & $30.0$ & RATCam g' & $17.29 \pm 0.04$ & $0.462 \pm 0.017$ \\
 & $77.17$ & $60.0$ & RATCam g' & $17.23 \pm 0.08$ & $0.489 \pm 0.036$ \\
 & $82.10$ & $60.0$ & RATCam g' & $17.36 \pm 0.03$ & $0.433 \pm 0.011$ \\
 & $87.05$ & $60.0$ & RATCam g' & $17.51 \pm 0.09$ & $0.378 \pm 0.032$ \\
 & $101.72$ & $60.0$ & RATCam g' & $17.61 \pm 0.07$ & $0.345 \pm 0.023$ \\
 & $106.55$ & $60.0$ & RATCam g' & $17.78 \pm 0.10$ & $0.295 \pm 0.026$ \\
 & $108.33$ & $120.0$ & RATCam g' & $17.76 \pm 0.05$ & $0.300 \pm 0.014$ \\
 & $116.27$ & $180.0$ & RATCam g' & $17.92 \pm 0.01$ & $0.257 \pm 0.003$ \\
 & $127.18$ & $120.0$ & RATCam g' & $18.13 \pm 0.04$ & $0.213 \pm 0.008$ \\
 & $135.08$ & $180.0$ & RATCam g' & $18.16 \pm 0.01$ & $0.207 \pm 0.002$ \\
 & $146.05$ & $120.0$ & RATCam g' & $18.34 \pm 0.02$ & $0.175 \pm 0.003$ \\
 & $153.98$ & $180.0$ & RATCam g' & $18.39 \pm 0.01$ & $0.168 \pm 0.001$ \\
 & $71.48$ & $30.0$ & RATCam i' & $16.48 \pm 0.09$ & $0.956 \pm 0.082$ \\
 & $74.85$ & $30.0$ & RATCam i' & $16.54 \pm 0.02$ & $0.896 \pm 0.013$ \\
 & $78.82$ & $60.0$ & RATCam i' & $16.55 \pm 0.01$ & $0.891 \pm 0.007$ \\
 & $83.80$ & $60.0$ & RATCam i' & $16.63 \pm 0.03$ & $0.831 \pm 0.024$ \\
 & $111.00$ & $120.0$ & RATCam i' & $17.09 \pm 0.04$ & $0.544 \pm 0.018$ \\
 & $119.93$ & $180.0$ & RATCam i' & $17.18 \pm 0.01$ & $0.498 \pm 0.006$ \\
 & $129.82$ & $120.0$ & RATCam i' & $17.32 \pm 0.04$ & $0.441 \pm 0.018$ \\
 & $138.77$ & $180.0$ & RATCam i' & $17.50 \pm 0.01$ & $0.371 \pm 0.004$ \\
 & $148.72$ & $120.0$ & RATCam i' & $17.57 \pm 0.01$ & $0.347 \pm 0.004$ \\
 & $6.82$ & $25.0$ & UVOT u & $18.75 \pm 0.39$ & $0.052 \pm 0.018$ \\
 & $7.23$ & $25.0$ & UVOT u & $18.71 \pm 0.37$ & $0.053 \pm 0.017$ \\
 & $7.65$ & $25.0$ & UVOT u & $18.58 \pm 0.35$ & $0.060 \pm 0.019$ \\
 & $8.07$ & $25.0$ & UVOT u & $18.33 \pm 0.30$ & $0.074 \pm 0.020$ \\
 & $8.48$ & $25.0$ & UVOT u & $17.92 \pm 0.22$ & $0.106 \pm 0.021$ \\
 & $8.90$ & $25.0$ & UVOT u & $17.32 \pm 0.16$ & $0.183 \pm 0.027$ \\
 & $9.32$ & $25.0$ & UVOT u & $17.06 \pm 0.14$ & $0.231 \pm 0.030$ \\
 & $12.23$ & $25.0$ & UVOT u & $15.58 \pm 0.08$ & $0.899 \pm 0.066$ \\
 & $18.90$ & $25.0$ & UVOT u & $14.97 \pm 0.07$ & $1.576 \pm 0.101$ \\
 & $19.32$ & $25.0$ & UVOT u & $14.88 \pm 0.14$ & $1.723 \pm 0.221$ \\
 & $21.82$ & $25.0$ & UVOT u & $15.15 \pm 0.07$ & $1.335 \pm 0.086$ \\
 & $22.23$ & $25.0$ & UVOT u & $14.98 \pm 0.15$ & $1.573 \pm 0.216$ \\
 & $24.73$ & $25.0$ & UVOT u & $15.30 \pm 0.07$ & $1.163 \pm 0.075$ \\
 & $27.65$ & $25.0$ & UVOT u & $15.53 \pm 0.08$ & $0.942 \pm 0.069$ \\
 & $30.57$ & $25.0$ & UVOT u & $15.67 \pm 0.09$ & $0.828 \pm 0.068$ \\
 & $9.78$ & $20.0$ & UVOT b & $17.08 \pm 0.12$ & $0.636 \pm 0.070$ \\
 & $12.45$ & $20.0$ & UVOT b & $15.80 \pm 0.10$ & $2.064 \pm 0.190$ \\
 & $12.78$ & $20.0$ & UVOT b & $15.77 \pm 0.08$ & $2.118 \pm 0.156$ \\
 & $19.12$ & $20.0$ & UVOT b & $15.32 \pm 0.13$ & $3.221 \pm 0.384$ \\
 & $19.45$ & $20.0$ & UVOT b & $15.22 \pm 0.06$ & $3.512 \pm 0.194$ \\
 & $22.12$ & $20.0$ & UVOT b & $15.41 \pm 0.10$ & $2.956 \pm 0.271$ \\
 & $22.45$ & $20.0$ & UVOT b & $15.47 \pm 0.07$ & $2.791 \pm 0.180$ \\
 & $25.12$ & $20.0$ & UVOT b & $15.77 \pm 0.08$ & $2.118 \pm 0.156$ \\
 & $25.45$ & $20.0$ & UVOT b & $15.72 \pm 0.11$ & $2.224 \pm 0.225$ \\
 & $28.12$ & $20.0$ & UVOT b & $15.92 \pm 0.07$ & $1.844 \pm 0.119$ \\
 & $30.78$ & $20.0$ & UVOT b & $16.01 \pm 0.12$ & $1.704 \pm 0.188$ \\
 & $31.12$ & $20.0$ & UVOT b & $16.11 \pm 0.09$ & $1.550 \pm 0.128$ \\
 & $10.92$ & $20.0$ & UVOT v & $15.88 \pm 0.14$ & $1.687 \pm 0.216$ \\
 & $11.25$ & $20.0$ & UVOT v & $15.63 \pm 0.18$ & $2.135 \pm 0.351$ \\
 & $13.92$ & $20.0$ & UVOT v & $15.04 \pm 0.08$ & $3.637 \pm 0.267$ \\
 & $17.58$ & $20.0$ & UVOT v & $14.62 \pm 0.09$ & $5.358 \pm 0.443$ \\
 & $17.92$ & $20.0$ & UVOT v & $14.65 \pm 0.09$ & $5.212 \pm 0.431$ \\
 & $20.58$ & $20.0$ & UVOT v & $14.65 \pm 0.08$ & $5.208 \pm 0.383$ \\
 & $20.92$ & $20.0$ & UVOT v & $14.80 \pm 0.12$ & $4.551 \pm 0.501$ \\
 & $23.58$ & $20.0$ & UVOT v & $15.19 \pm 0.08$ & $3.167 \pm 0.233$ \\
 & $26.25$ & $20.0$ & UVOT v & $15.10 \pm 0.13$ & $3.456 \pm 0.412$ \\
 & $26.58$ & $20.0$ & UVOT v & $15.26 \pm 0.10$ & $2.974 \pm 0.273$ \\
 & $29.25$ & $20.0$ & UVOT v & $15.24 \pm 0.10$ & $3.029 \pm 0.278$ \\
 & $29.58$ & $20.0$ & UVOT v & $15.52 \pm 0.17$ & $2.359 \pm 0.366$ \\
110726A & $3.19$ & $72.0$ & RINGO r' & $17.67 \pm 0.07$ & $0.367 \pm 0.023$ \\
 & $4.41$ & $72.0$ & RINGO r' & $17.97 \pm 0.09$ & $0.278 \pm 0.023$ \\
 & $5.65$ & $72.0$ & RINGO r' & $18.13 \pm 0.10$ & $0.241 \pm 0.022$ \\
 & $6.89$ & $72.0$ & RINGO r' & $18.33 \pm 0.12$ & $0.200 \pm 0.022$ \\
 & $8.13$ & $72.0$ & RINGO r' & $18.56 \pm 0.14$ & $0.163 \pm 0.020$ \\
 & $9.37$ & $72.0$ & RINGO r' & $18.63 \pm 0.14$ & $0.152 \pm 0.020$ \\
 & $10.61$ & $72.0$ & RINGO r' & $18.79 \pm 0.18$ & $0.133 \pm 0.021$ \\
 & $11.85$ & $72.0$ & RINGO r' & $18.74 \pm 0.16$ & $0.138 \pm 0.020$ \\
 & $14.73$ & $30.0$ & RATCam r' & $19.25 \pm 0.07$ & $0.086 \pm 0.006$ \\
 & $16.38$ & $30.0$ & RATCam r' & $19.34 \pm 0.08$ & $0.079 \pm 0.006$ \\
 & $20.53$ & $30.0$ & RATCam r' & $19.56 \pm 0.07$ & $0.065 \pm 0.004$ \\
 & $24.20$ & $30.0$ & RATCam r' & $19.82 \pm 0.09$ & $0.051 \pm 0.004$ \\
 & $28.81$ & $60.0$ & RATCam r' & $19.78 \pm 0.06$ & $0.053 \pm 0.003$ \\
 & $34.11$ & $60.0$ & RATCam r' & $19.86 \pm 0.06$ & $0.049 \pm 0.003$ \\
 & $39.53$ & $60.0$ & RATCam r' & $19.89 \pm 0.06$ & $0.047 \pm 0.003$ \\
 & $45.02$ & $60.0$ & RATCam r' & $19.90 \pm 0.05$ & $0.047 \pm 0.002$ \\
 & $50.47$ & $60.0$ & RATCam r' & $20.06 \pm 0.07$ & $0.041 \pm 0.003$ \\
 & $55.87$ & $60.0$ & RATCam r' & $19.98 \pm 0.06$ & $0.044 \pm 0.002$ \\
 & $87.59$ & $1800.0$ & RATCam r' & $20.62 \pm 0.03$ & $0.024 \pm 0.001$ \\
 & $136.32$ & $1800.0$ & RATCam r' & $21.14 \pm 0.05$ & $0.015 \pm 0.001$ \\
 & $18.16$ & $30.0$ & RATCam g' & $19.70 \pm 0.06$ & $0.061 \pm 0.003$ \\
 & $21.75$ & $30.0$ & RATCam g' & $19.91 \pm 0.07$ & $0.050 \pm 0.003$ \\
 & $25.40$ & $60.0$ & RATCam g' & $20.06 \pm 0.05$ & $0.044 \pm 0.002$ \\
 & $30.55$ & $60.0$ & RATCam g' & $20.15 \pm 0.05$ & $0.040 \pm 0.002$ \\
 & $35.91$ & $60.0$ & RATCam g' & $20.22 \pm 0.05$ & $0.038 \pm 0.002$ \\
 & $41.35$ & $60.0$ & RATCam g' & $20.29 \pm 0.05$ & $0.035 \pm 0.002$ \\
 & $48.41$ & $60.0$ & RATCam g' & $20.37 \pm 0.06$ & $0.033 \pm 0.002$ \\
 & $52.26$ & $60.0$ & RATCam g' & $20.48 \pm 0.06$ & $0.030 \pm 0.002$ \\
 & $57.68$ & $60.0$ & RATCam g' & $20.55 \pm 0.07$ & $0.028 \pm 0.002$ \\
 & $19.37$ & $30.0$ & RATCam i' & $19.22 \pm 0.09$ & $0.085 \pm 0.007$ \\
 & $23.00$ & $30.0$ & RATCam i' & $19.40 \pm 0.10$ & $0.072 \pm 0.007$ \\
 & $27.10$ & $60.0$ & RATCam i' & $19.47 \pm 0.08$ & $0.067 \pm 0.005$ \\
 & $32.25$ & $60.0$ & RATCam i' & $19.72 \pm 0.09$ & $0.053 \pm 0.004$ \\
 & $37.62$ & $60.0$ & RATCam i' & $19.61 \pm 0.08$ & $0.059 \pm 0.004$ \\
 & $43.12$ & $60.0$ & RATCam i' & $19.72 \pm 0.10$ & $0.053 \pm 0.005$ \\
 & $48.56$ & $60.0$ & RATCam i' & $19.73 \pm 0.09$ & $0.053 \pm 0.004$ \\
 & $54.02$ & $60.0$ & RATCam i' & $19.57 \pm 0.08$ & $0.061 \pm 0.004$ \\
 & $59.41$ & $60.0$ & RATCam i' & $19.77 \pm 0.10$ & $0.051 \pm 0.005$ \\
 & $1350.49$ & $7200.0$ & RATCam i' & $23.03 \pm 0.29$ & $0.003 \pm 0.001$ \\
120119A & $3.24$ & $60.2$ & RINGO r' & $17.77 \pm 0.05$ & $0.356 \pm 0.017$ \\
 & $4.24$ & $60.2$ & RINGO r' & $17.64 \pm 0.05$ & $0.401 \pm 0.019$ \\
 & $5.25$ & $60.2$ & RINGO r' & $17.57 \pm 0.04$ & $0.427 \pm 0.016$ \\
 & $6.25$ & $59.4$ & RINGO r' & $17.50 \pm 0.06$ & $0.456 \pm 0.026$ \\
 & $7.24$ & $60.2$ & RINGO r' & $17.57 \pm 0.03$ & $0.427 \pm 0.012$ \\
 & $8.24$ & $60.2$ & RINGO r' & $17.55 \pm 0.04$ & $0.435 \pm 0.017$ \\
 & $9.25$ & $59.4$ & RINGO r' & $17.34 \pm 0.02$ & $0.528 \pm 0.011$ \\
 & $10.24$ & $60.2$ & RINGO r' & $17.26 \pm 0.03$ & $0.568 \pm 0.017$ \\
 & $11.24$ & $60.2$ & RINGO r' & $17.24 \pm 0.03$ & $0.579 \pm 0.017$ \\
 & $12.24$ & $58.9$ & RINGO r' & $17.26 \pm 0.03$ & $0.568 \pm 0.017$ \\
 & $14.55$ & $10.0$ & RATCam r' & $17.32 \pm 0.04$ & $0.538 \pm 0.020$ \\
 & $14.91$ & $10.0$ & RATCam r' & $17.40 \pm 0.05$ & $0.500 \pm 0.023$ \\
 & $15.28$ & $10.0$ & RATCam r' & $17.41 \pm 0.04$ & $0.495 \pm 0.018$ \\
 & $16.22$ & $20.0$ & RATCam r' & $17.46 \pm 0.04$ & $0.473 \pm 0.017$ \\
 & $17.64$ & $120.0$ & RATCam r' & $17.59 \pm 0.01$ & $0.419 \pm 0.004$ \\
 & $25.68$ & $120.0$ & RATCam r' & $17.94 \pm 0.02$ & $0.304 \pm 0.006$ \\
 & $27.90$ & $120.0$ & RATCam r' & $18.01 \pm 0.02$ & $0.285 \pm 0.005$ \\
 & $40.93$ & $120.0$ & RATCam r' & $18.43 \pm 0.03$ & $0.193 \pm 0.005$ \\
 & $43.12$ & $120.0$ & RATCam r' & $18.54 \pm 0.03$ & $0.175 \pm 0.005$ \\
 & $45.33$ & $120.0$ & RATCam r' & $18.72 \pm 0.04$ & $0.148 \pm 0.005$ \\
 & $20.27$ & $120.0$ & RATCam i' & $16.81 \pm 0.01$ & $0.811 \pm 0.007$ \\
 & $30.57$ & $120.0$ & RATCam i' & $17.24 \pm 0.01$ & $0.546 \pm 0.005$ \\
 & $32.77$ & $120.0$ & RATCam i' & $17.33 \pm 0.02$ & $0.503 \pm 0.009$ \\
 & $48.21$ & $120.0$ & RATCam i' & $18.01 \pm 0.03$ & $0.269 \pm 0.007$ \\
 & $50.41$ & $120.0$ & RATCam i' & $18.09 \pm 0.03$ & $0.250 \pm 0.007$ \\
 & $52.62$ & $120.0$ & RATCam i' & $18.11 \pm 0.04$ & $0.245 \pm 0.009$ \\
 & $22.89$ & $120.0$ & RATCam z' & $16.38 \pm 0.01$ & $1.155 \pm 0.011$ \\
 & $35.47$ & $120.0$ & RATCam z' & $16.68 \pm 0.02$ & $0.876 \pm 0.016$ \\
 & $37.67$ & $120.0$ & RATCam z' & $16.75 \pm 0.02$ & $0.821 \pm 0.015$ \\
 & $1.97$ & $20.0$ & PROMPT R & $17.82 \pm 0.13$ & $0.269 \pm 0.032$ \\
 & $2.45$ & $20.0$ & PROMPT R & $17.70 \pm 0.14$ & $0.300 \pm 0.039$ \\
 & $2.94$ & $20.0$ & PROMPT R & $17.55 \pm 0.11$ & $0.344 \pm 0.035$ \\
 & $3.41$ & $20.0$ & PROMPT R & $17.41 \pm 0.10$ & $0.391 \pm 0.036$ \\
 & $3.90$ & $20.0$ & PROMPT R & $17.55 \pm 0.11$ & $0.344 \pm 0.035$ \\
 & $4.81$ & $40.0$ & PROMPT R & $17.36 \pm 0.06$ & $0.408 \pm 0.023$ \\
 & $5.64$ & $40.0$ & PROMPT R & $17.49 \pm 0.06$ & $0.362 \pm 0.020$ \\
 & $6.48$ & $40.0$ & PROMPT R & $17.38 \pm 0.05$ & $0.400 \pm 0.018$ \\
 & $7.65$ & $80.0$ & PROMPT R & $17.30 \pm 0.03$ & $0.431 \pm 0.012$ \\
 & $9.19$ & $80.0$ & PROMPT R & $17.18 \pm 0.03$ & $0.481 \pm 0.013$ \\
 & $20.16$ & $160.0$ & PROMPT R & $17.34 \pm 0.03$ & $0.415 \pm 0.011$ \\
 & $23.18$ & $160.0$ & PROMPT R & $17.39 \pm 0.03$ & $0.397 \pm 0.011$ \\
 & $26.18$ & $160.0$ & PROMPT R & $17.62 \pm 0.03$ & $0.321 \pm 0.009$ \\
 & $29.20$ & $160.0$ & PROMPT R & $17.70 \pm 0.03$ & $0.298 \pm 0.008$ \\
 & $32.93$ & $240.0$ & PROMPT R & $17.83 \pm 0.03$ & $0.264 \pm 0.007$ \\
 & $38.19$ & $320.0$ & PROMPT R & $18.03 \pm 0.03$ & $0.220 \pm 0.006$ \\
 & $59.16$ & $560.0$ & PROMPT R & $18.77 \pm 0.04$ & $0.111 \pm 0.004$ \\
 & $69.57$ & $560.0$ & PROMPT R & $18.90 \pm 0.04$ & $0.099 \pm 0.004$ \\
 & $81.29$ & $560.0$ & PROMPT R & $19.16 \pm 0.06$ & $0.078 \pm 0.004$ \\
 & $94.00$ & $640.0$ & PROMPT R & $19.53 \pm 0.08$ & $0.055 \pm 0.004$ \\
 & $108.24$ & $800.0$ & PROMPT R & $19.68 \pm 0.08$ & $0.048 \pm 0.004$ \\
 & $123.06$ & $720.0$ & PROMPT R & $19.89 \pm 0.11$ & $0.040 \pm 0.004$ \\
 & $143.61$ & $880.0$ & PROMPT R & $20.04 \pm 0.12$ & $0.035 \pm 0.004$ \\
 & $165.14$ & $1040.0$ & PROMPT R & $20.18 \pm 0.15$ & $0.031 \pm 0.004$ \\
 & $186.06$ & $1040.0$ & PROMPT R & $20.29 \pm 0.16$ & $0.028 \pm 0.004$ \\
 & $208.34$ & $1040.0$ & PROMPT R & $20.55 \pm 0.18$ & $0.022 \pm 0.004$ \\
 & $240.71$ & $2080.0$ & PROMPT R & $20.64 \pm 0.14$ & $0.020 \pm 0.003$ \\
 & $278.12$ & $1520.0$ & PROMPT R & $20.67 \pm 0.20$ & $0.020 \pm 0.004$ \\
 & $1.97$ & $20.0$ & PROMPT I & $16.29 \pm 0.06$ & $0.808 \pm 0.045$ \\
 & $2.45$ & $20.0$ & PROMPT I & $16.39 \pm 0.07$ & $0.737 \pm 0.047$ \\
 & $2.95$ & $20.0$ & PROMPT I & $16.23 \pm 0.06$ & $0.854 \pm 0.047$ \\
 & $3.37$ & $10.0$ & PROMPT I & $16.25 \pm 0.11$ & $0.841 \pm 0.085$ \\
 & $3.99$ & $40.0$ & PROMPT I & $16.21 \pm 0.04$ & $0.869 \pm 0.032$ \\
 & $4.81$ & $40.0$ & PROMPT I & $16.17 \pm 0.04$ & $0.902 \pm 0.033$ \\
 & $5.64$ & $40.0$ & PROMPT I & $16.11 \pm 0.04$ & $0.953 \pm 0.035$ \\
 & $6.48$ & $40.0$ & PROMPT I & $16.14 \pm 0.04$ & $0.927 \pm 0.034$ \\
 & $7.65$ & $80.0$ & PROMPT I & $16.08 \pm 0.03$ & $0.979 \pm 0.027$ \\
 & $9.17$ & $80.0$ & PROMPT I & $15.99 \pm 0.02$ & $1.064 \pm 0.020$ \\
 & $20.16$ & $160.0$ & PROMPT I & $16.12 \pm 0.03$ & $0.944 \pm 0.026$ \\
 & $23.18$ & $160.0$ & PROMPT I & $16.24 \pm 0.02$ & $0.845 \pm 0.016$ \\
 & $26.19$ & $160.0$ & PROMPT I & $16.33 \pm 0.03$ & $0.778 \pm 0.021$ \\
 & $29.22$ & $160.0$ & PROMPT I & $16.46 \pm 0.03$ & $0.690 \pm 0.019$ \\
 & $32.96$ & $240.0$ & PROMPT I & $16.63 \pm 0.03$ & $0.590 \pm 0.016$ \\
 & $38.97$ & $240.0$ & PROMPT I & $16.88 \pm 0.03$ & $0.469 \pm 0.013$ \\
 & $58.72$ & $560.0$ & PROMPT I & $17.43 \pm 0.04$ & $0.283 \pm 0.010$ \\
 & $69.64$ & $560.0$ & PROMPT I & $17.75 \pm 0.04$ & $0.210 \pm 0.008$ \\
 & $81.33$ & $560.0$ & PROMPT I & $17.95 \pm 0.05$ & $0.175 \pm 0.008$ \\
 & $94.02$ & $640.0$ & PROMPT I & $18.26 \pm 0.06$ & $0.132 \pm 0.007$ \\
 & $108.30$ & $800.0$ & PROMPT I & $18.43 \pm 0.06$ & $0.113 \pm 0.006$ \\
 & $123.96$ & $800.0$ & PROMPT I & $18.64 \pm 0.08$ & $0.093 \pm 0.007$ \\
 & $143.15$ & $800.0$ & PROMPT I & $18.83 \pm 0.09$ & $0.078 \pm 0.006$ \\
 & $163.97$ & $880.0$ & PROMPT I & $18.95 \pm 0.12$ & $0.070 \pm 0.008$ \\
 & $186.98$ & $800.0$ & PROMPT I & $19.18 \pm 0.14$ & $0.057 \pm 0.007$ \\
 & $209.23$ & $800.0$ & PROMPT I & $19.39 \pm 0.17$ & $0.047 \pm 0.007$ \\
 & $242.54$ & $1360.0$ & PROMPT I & $19.35 \pm 0.11$ & $0.048 \pm 0.005$ \\
 & $280.31$ & $1040.0$ & PROMPT I & $20.01 \pm 0.32$ & $0.027 \pm 0.008$ \\
120308A & $4.00$ & $24.4$ & RINGO R & $16.14 \pm 0.06$ & $1.281 \pm 0.071$ \\
 & $4.40$ & $25.2$ & RINGO R & $15.83 \pm 0.06$ & $1.704 \pm 0.094$ \\
 & $4.82$ & $25.2$ & RINGO R & $15.83 \pm 0.06$ & $1.704 \pm 0.094$ \\
 & $5.24$ & $25.2$ & RINGO R & $15.84 \pm 0.06$ & $1.688 \pm 0.093$ \\
 & $5.66$ & $25.2$ & RINGO R & $15.94 \pm 0.06$ & $1.540 \pm 0.085$ \\
 & $6.08$ & $25.2$ & RINGO R & $16.05 \pm 0.06$ & $1.391 \pm 0.077$ \\
 & $6.50$ & $25.2$ & RINGO R & $16.21 \pm 0.06$ & $1.201 \pm 0.066$ \\
 & $6.92$ & $25.2$ & RINGO R & $16.12 \pm 0.06$ & $1.304 \pm 0.072$ \\
 & $7.34$ & $25.2$ & RINGO R & $16.27 \pm 0.06$ & $1.136 \pm 0.063$ \\
 & $7.76$ & $25.2$ & RINGO R & $16.30 \pm 0.06$ & $1.105 \pm 0.061$ \\
 & $8.18$ & $25.2$ & RINGO R & $16.34 \pm 0.06$ & $1.065 \pm 0.059$ \\
 & $8.60$ & $25.2$ & RINGO R & $16.35 \pm 0.06$ & $1.055 \pm 0.058$ \\
 & $9.02$ & $25.2$ & RINGO R & $16.50 \pm 0.07$ & $0.920 \pm 0.059$ \\
 & $0.00$ & $0.0$ & RINGO R & $0.00 \pm 0.00$ & $0.000 \pm 0.000$ \\
 & $9.58$ & $83.9$ & RINGO R & $16.63 \pm 0.06$ & $0.816 \pm 0.045$ \\
 & $10.98$ & $84.0$ & RINGO R & $16.78 \pm 0.06$ & $0.710 \pm 0.039$ \\
 & $12.38$ & $84.0$ & RINGO R & $16.87 \pm 0.07$ & $0.654 \pm 0.042$ \\
 & $0.00$ & $0.0$ & RINGO R & $0.00 \pm 0.00$ & $0.000 \pm 0.000$ \\
 & $15.84$ & $30.0$ & RATCam r' & $17.15 \pm 0.09$ & $0.548 \pm 0.045$ \\
 & $17.32$ & $30.0$ & RATCam r' & $17.10 \pm 0.09$ & $0.574 \pm 0.047$ \\
 & $18.55$ & $120.0$ & RATCam r' & $17.19 \pm 0.08$ & $0.528 \pm 0.039$ \\
 & $26.18$ & $120.0$ & RATCam r' & $17.46 \pm 0.09$ & $0.412 \pm 0.034$ \\
 & $28.38$ & $120.0$ & RATCam r' & $17.63 \pm 0.09$ & $0.352 \pm 0.029$ \\
 & $40.53$ & $120.0$ & RATCam r' & $18.17 \pm 0.22$ & $0.218 \pm 0.044$ \\
 & $21.13$ & $120.0$ & RATCam i' & $17.01 \pm 0.04$ & $0.606 \pm 0.022$ \\
 & $31.02$ & $120.0$ & RATCam i' & $17.39 \pm 0.06$ & $0.428 \pm 0.024$ \\
 & $33.22$ & $120.0$ & RATCam i' & $17.50 \pm 0.09$ & $0.387 \pm 0.032$ \\
 & $23.73$ & $120.0$ & RATCam z' & $16.77 \pm 0.18$ & $0.754 \pm 0.124$ \\
 & $36.93$ & $240.0$ & RATCam z' & $17.11 \pm 0.19$ & $0.552 \pm 0.096$ \\
120311A & $3.01$ & $60.5$ & RINGO r' & $17.90 \pm 0.11$ & $0.335 \pm 0.035$ \\
 & $4.02$ & $59.7$ & RINGO r' & $18.10 \pm 0.13$ & $0.279 \pm 0.034$ \\
 & $5.01$ & $60.5$ & RINGO r' & $18.44 \pm 0.16$ & $0.205 \pm 0.030$ \\
 & $6.02$ & $119.4$ & RINGO r' & $18.96 \pm 0.17$ & $0.127 \pm 0.019$ \\
 & $8.01$ & $60.5$ & RINGO r' & $18.50 \pm 0.16$ & $0.194 \pm 0.028$ \\
 & $9.02$ & $119.4$ & RINGO r' & $19.22 \pm 0.20$ & $0.101 \pm 0.018$ \\
 & $11.01$ & $118.8$ & RINGO r' & $19.77 \pm 0.35$ & $0.063 \pm 0.020$ \\
 & $14.51$ & $30.0$ & RATCam r' & $19.66 \pm 0.17$ & $0.067 \pm 0.010$ \\
 & $16.05$ & $30.0$ & RATCam r' & $19.90 \pm 0.21$ & $0.054 \pm 0.010$ \\
 & $17.56$ & $120.0$ & RATCam r' & $19.83 \pm 0.11$ & $0.057 \pm 0.006$ \\
 & $27.71$ & $240.0$ & RATCam r' & $20.26 \pm 0.11$ & $0.038 \pm 0.004$ \\
 & $44.99$ & $360.0$ & RATCam r' & $20.85 \pm 0.13$ & $0.022 \pm 0.003$ \\
 & $21.16$ & $120.0$ & RATCam i' & $19.67 \pm 0.13$ & $0.061 \pm 0.007$ \\
 & $32.84$ & $240.0$ & RATCam i' & $20.08 \pm 0.13$ & $0.042 \pm 0.005$ \\
 & $54.27$ & $360.0$ & RATCam i' & $20.92 \pm 0.28$ & $0.020 \pm 0.005$ \\
 & $25.10$ & $120.0$ & RATCam z' & $19.57 \pm 0.17$ & $0.064 \pm 0.010$ \\
 & $38.65$ & $240.0$ & RATCam z' & $20.45 \pm 0.27$ & $0.029 \pm 0.007$ \\
120326A & $3.60$ & $240.0$ & RINGO r' & $18.52 \pm 0.18$ & $0.135 \pm 0.022$ \\
 & $9.20$ & $320.0$ & RINGO r' & $18.74 \pm 0.18$ & $0.110 \pm 0.018$ \\
 & $14.02$ & $30.0$ & RATCam r' & $18.81 \pm 0.16$ & $0.103 \pm 0.015$ \\
 & $15.52$ & $30.0$ & RATCam r' & $19.00 \pm 0.14$ & $0.086 \pm 0.011$ \\
 & $16.75$ & $120.0$ & RATCam r' & $18.95 \pm 0.04$ & $0.090 \pm 0.003$ \\
 & $24.63$ & $120.0$ & RATCam r' & $19.25 \pm 0.05$ & $0.068 \pm 0.003$ \\
 & $26.83$ & $120.0$ & RATCam r' & $19.33 \pm 0.06$ & $0.063 \pm 0.003$ \\
 & $39.38$ & $120.0$ & RATCam r' & $19.40 \pm 0.06$ & $0.059 \pm 0.003$ \\
 & $41.58$ & $120.0$ & RATCam r' & $19.40 \pm 0.05$ & $0.059 \pm 0.003$ \\
 & $43.78$ & $120.0$ & RATCam r' & $19.38 \pm 0.06$ & $0.060 \pm 0.003$ \\
 & $19.42$ & $120.0$ & RATCam i' & $19.06 \pm 0.07$ & $0.094 \pm 0.006$ \\
 & $29.58$ & $120.0$ & RATCam i' & $19.31 \pm 0.06$ & $0.074 \pm 0.004$ \\
 & $46.55$ & $360.0$ & RATCam i' & $19.40 \pm 0.07$ & $0.068 \pm 0.004$ \\
 & $22.10$ & $120.0$ & RATCam z' & $18.79 \pm 0.13$ & $0.118 \pm 0.014$ \\
 & $34.58$ & $240.0$ & RATCam z' & $18.87 \pm 0.10$ & $0.110 \pm 0.010$ \\
 & $54.02$ & $360.0$ & RATCam z' & $19.05 \pm 0.10$ & $0.093 \pm 0.009$ \\
120327A & $27.73$ & $60.5$ & RINGO r' & $16.80 \pm 0.06$ & $1.403 \pm 0.075$ \\
 & $28.74$ & $59.6$ & RINGO r' & $16.85 \pm 0.06$ & $1.340 \pm 0.072$ \\
 & $29.74$ & $60.5$ & RINGO r' & $16.97 \pm 0.05$ & $1.199 \pm 0.059$ \\
 & $30.74$ & $59.6$ & RINGO r' & $17.00 \pm 0.05$ & $1.167 \pm 0.058$ \\
 & $31.74$ & $60.5$ & RINGO r' & $17.09 \pm 0.05$ & $1.074 \pm 0.053$ \\
 & $32.75$ & $59.6$ & RINGO r' & $17.08 \pm 0.06$ & $1.084 \pm 0.064$ \\
 & $33.74$ & $59.6$ & RINGO r' & $17.13 \pm 0.05$ & $1.035 \pm 0.051$ \\
 & $34.73$ & $60.5$ & RINGO r' & $17.07 \pm 0.06$ & $1.094 \pm 0.059$ \\
 & $35.74$ & $59.6$ & RINGO r' & $17.16 \pm 0.05$ & $1.007 \pm 0.050$ \\
 & $36.73$ & $58.9$ & RINGO r' & $17.19 \pm 0.06$ & $0.980 \pm 0.058$ \\
 & $43.42$ & $60.4$ & RINGO r' & $17.41 \pm 0.05$ & $0.800 \pm 0.040$ \\
 & $44.43$ & $59.6$ & RINGO r' & $17.44 \pm 0.06$ & $0.778 \pm 0.042$ \\
 & $45.42$ & $59.2$ & RINGO r' & $17.52 \pm 0.06$ & $0.723 \pm 0.039$ \\
 & $39.22$ & $30.0$ & RATCam r' & $17.35 \pm 0.07$ & $0.846 \pm 0.055$ \\
 & $40.81$ & $30.0$ & RATCam r' & $17.39 \pm 0.07$ & $0.815 \pm 0.053$ \\
 & $50.15$ & $30.0$ & RATCam r' & $17.65 \pm 0.06$ & $0.641 \pm 0.038$ \\
 & $53.63$ & $30.0$ & RATCam r' & $17.72 \pm 0.07$ & $0.602 \pm 0.039$ \\
 & $58.22$ & $60.0$ & RATCam r' & $17.86 \pm 0.06$ & $0.529 \pm 0.028$ \\
 & $63.47$ & $60.0$ & RATCam r' & $17.95 \pm 0.06$ & $0.486 \pm 0.026$ \\
 & $68.72$ & $60.0$ & RATCam r' & $18.10 \pm 0.06$ & $0.424 \pm 0.025$ \\
 & $74.23$ & $60.0$ & RATCam r' & $18.19 \pm 0.05$ & $0.390 \pm 0.019$ \\
 & $81.02$ & $60.0$ & RATCam r' & $18.31 \pm 0.05$ & $0.349 \pm 0.017$ \\
 & $93.72$ & $120.0$ & RATCam r' & $18.49 \pm 0.05$ & $0.296 \pm 0.015$ \\
 & $106.53$ & $180.0$ & RATCam r' & $18.68 \pm 0.05$ & $0.248 \pm 0.012$ \\
 & $121.60$ & $120.0$ & RATCam r' & $18.83 \pm 0.06$ & $0.216 \pm 0.012$ \\
 & $136.05$ & $180.0$ & RATCam r' & $18.97 \pm 0.05$ & $0.190 \pm 0.009$ \\
 & $160.27$ & $600.0$ & RATCam r' & $19.21 \pm 0.05$ & $0.152 \pm 0.008$ \\
 & $47.88$ & $30.0$ & RATCam g' & $18.72 \pm 0.11$ & $0.329 \pm 0.034$ \\
 & $51.32$ & $30.0$ & RATCam g' & $18.78 \pm 0.12$ & $0.312 \pm 0.035$ \\
 & $54.82$ & $60.0$ & RATCam g' & $18.78 \pm 0.09$ & $0.311 \pm 0.025$ \\
 & $59.95$ & $60.0$ & RATCam g' & $18.93 \pm 0.09$ & $0.271 \pm 0.023$ \\
 & $65.25$ & $60.0$ & RATCam g' & $19.05 \pm 0.10$ & $0.242 \pm 0.022$ \\
 & $70.45$ & $60.0$ & RATCam g' & $19.18 \pm 0.10$ & $0.215 \pm 0.020$ \\
 & $76.45$ & $60.0$ & RATCam g' & $19.35 \pm 0.11$ & $0.184 \pm 0.018$ \\
 & $82.87$ & $60.0$ & RATCam g' & $19.48 \pm 0.09$ & $0.163 \pm 0.013$ \\
 & $86.85$ & $120.0$ & RATCam g' & $19.52 \pm 0.09$ & $0.157 \pm 0.012$ \\
 & $97.75$ & $180.0$ & RATCam g' & $19.63 \pm 0.08$ & $0.142 \pm 0.010$ \\
 & $112.20$ & $120.0$ & RATCam g' & $19.81 \pm 0.09$ & $0.120 \pm 0.010$ \\
 & $125.75$ & $180.0$ & RATCam g' & $19.90 \pm 0.08$ & $0.111 \pm 0.008$ \\
 & $140.67$ & $120.0$ & RATCam g' & $20.05 \pm 0.08$ & $0.096 \pm 0.007$ \\
 & $177.99$ & $600.0$ & RATCam g' & $20.39 \pm 0.08$ & $0.070 \pm 0.005$ \\
 & $49.02$ & $30.0$ & RATCam i' & $17.16 \pm 0.04$ & $0.839 \pm 0.033$ \\
 & $52.45$ & $30.0$ & RATCam i' & $17.22 \pm 0.04$ & $0.794 \pm 0.031$ \\
 & $56.48$ & $60.0$ & RATCam i' & $17.31 \pm 0.04$ & $0.731 \pm 0.024$ \\
 & $61.62$ & $60.0$ & RATCam i' & $17.47 \pm 0.04$ & $0.631 \pm 0.025$ \\
 & $66.97$ & $60.0$ & RATCam i' & $17.56 \pm 0.04$ & $0.581 \pm 0.023$ \\
 & $72.13$ & $60.0$ & RATCam i' & $17.72 \pm 0.04$ & $0.501 \pm 0.020$ \\
 & $78.15$ & $60.0$ & RATCam i' & $17.82 \pm 0.04$ & $0.457 \pm 0.015$ \\
 & $84.60$ & $60.0$ & RATCam i' & $17.92 \pm 0.05$ & $0.417 \pm 0.019$ \\
 & $89.57$ & $120.0$ & RATCam i' & $18.05 \pm 0.04$ & $0.370 \pm 0.012$ \\
 & $101.75$ & $180.0$ & RATCam i' & $18.18 \pm 0.04$ & $0.328 \pm 0.011$ \\
 & $114.95$ & $120.0$ & RATCam i' & $18.29 \pm 0.04$ & $0.296 \pm 0.012$ \\
 & $129.65$ & $180.0$ & RATCam i' & $18.39 \pm 0.04$ & $0.270 \pm 0.009$ \\
 & $143.57$ & $120.0$ & RATCam i' & $18.59 \pm 0.04$ & $0.225 \pm 0.009$ \\
\end{longtable}

\newpage

\bibliographystyle{yahapj}
\bibliography{references}

\begin{thebibliography}{}
\providecommand\natexlab[1]{#1}
\providecommand\JournalTitle[1]{#1}

\bibitem[{{Arnold}(2017)}]{arnold}
{Arnold}, D.~M. 2017, \JournalTitle{PhD Thesis, Liverpool JMU}

\bibitem[{{Bersier}(2012)}]{gcn13320}
{Bersier}, D. 2012, \JournalTitle{GRB Coordinates Network}, 13320

\bibitem[{{Bertin} \& {Arnouts}(1996)}]{sextractor}
{Bertin}, E., \& {Arnouts}, S. 1996,
  \href{http://dx.doi.org/10.1051/aas:1996164}{\JournalTitle{\aaps}, 117, 393}

\bibitem[{{Beuermann} {et~al.}(1999){Beuermann}, {Hessman}, {Reinsch},
  {Nicklas}, {Vreeswijk}, {Galama}, {Rol}, {van Paradijs}, {Kouveliotou},
  {Frontera}, {Masetti}, {Palazzi}, \& {Pian}}]{beuermann1999}
{Beuermann}, K., {Hessman}, F.~V., {Reinsch}, K., {et~al.} 1999,
  \JournalTitle{\aap}, 352, L26

\bibitem[{{Breeveld} {et~al.}(2011){Breeveld}, {Landsman}, {Holland}, {Roming},
  {Kuin}, \& {Page}}]{breeveld}
{Breeveld}, A.~A., {Landsman}, W., {Holland}, S.~T., {et~al.} 2011,
  \href{http://dx.doi.org/10.1063/1.3621807}{in American Institute of Physics
  Conference Series, Vol. 1358, American Institute of Physics Conference
  Series, ed. J.~E. {McEnery}, J.~L. {Racusin}, \& N.~{Gehrels}}, 373

\bibitem[{{Cenko} {et~al.}(2011){Cenko}, {Hora}, \& {Bloom}}]{gcn11638}
{Cenko}, S.~B., {Hora}, J.~L., \& {Bloom}, J.~S. 2011, \JournalTitle{GRB
  Coordinates Network}, 11638

\bibitem[{{Clarke} \& {Neumayer}(2002)}]{clarke}
{Clarke}, D., \& {Neumayer}, D. 2002,
  \href{http://dx.doi.org/10.1051/0004-6361:20011717}{\JournalTitle{\aap}, 383,
  360}

\bibitem[{{Covino} {et~al.}(1999){Covino}, {Lazzati}, {Ghisellini}, {Saracco},
  {Campana}, {Chincarini}, {di Serego}, {Cimatti}, {Vanzi}, {Pasquini},
  {Haardt}, {Israel}, {Stella}, \& {Vietri}}]{covino1999}
{Covino}, S., {Lazzati}, D., {Ghisellini}, G., {et~al.} 1999,
  \JournalTitle{\aap}, 348, L1

\bibitem[{{Cox}(1976)}]{cox}
{Cox}, L.~J. 1976,
  \href{http://dx.doi.org/10.1093/mnras/176.3.525}{\JournalTitle{\mnras}, 176,
  525}

\bibitem[{{Cucchiara} {et~al.}(2011{\natexlab{a}}){Cucchiara}, {Bloom}, \&
  {Cenko}}]{gcn12202}
{Cucchiara}, A., {Bloom}, J.~S., \& {Cenko}, S.~B. 2011{\natexlab{a}},
  \JournalTitle{GRB Coordinates Network}, 12202

\bibitem[{{Cucchiara} \& {Prochaska}(2012)}]{gcn12865}
{Cucchiara}, A., \& {Prochaska}, J.~X. 2012, \JournalTitle{GRB Coordinates
  Network}, 12865

\bibitem[{{Cucchiara} {et~al.}(2011{\natexlab{b}}){Cucchiara}, {Cenko},
  {Bloom}, {Melandri}, {Morgan}, {Kobayashi}, {Smith}, {Perley}, {Li}, {Hora},
  {da Silva}, {Prochaska}, {Milne}, {Butler}, {Cobb}, {Worseck}, {Mundell},
  {Steele}, {Filippenko}, {Fumagalli}, {Klein}, {Stephens}, {Bluck}, \&
  {Mason}}]{cucchiara}
{Cucchiara}, A., {Cenko}, S.~B., {Bloom}, J.~S., {et~al.} 2011{\natexlab{b}},
  \href{http://dx.doi.org/10.1088/0004-637X/743/2/154}{\JournalTitle{\apj},
  743, 154}

\bibitem[{{D'Avanzo} {et~al.}(2012){D'Avanzo}, {Milvang-Jensen}, {Vreeswijk},
  {Tanvir}, {Malesani}, {Goldoni}, {Kaper}, {Covino}, \& {Fynbo}}]{gcn13051}
{D'Avanzo}, P., {Milvang-Jensen}, B., {Vreeswijk}, P., {et~al.} 2012,
  \JournalTitle{GRB Coordinates Network}, 13051

\bibitem[{{Fukugita} {et~al.}(1995){Fukugita}, {Shimasaku}, \&
  {Ichikawa}}]{fuk}
{Fukugita}, M., {Shimasaku}, K., \& {Ichikawa}, T. 1995,
  \href{http://dx.doi.org/10.1086/133643}{\JournalTitle{\pasp}, 107, 945}

\bibitem[{{Ghisellini} \& {Lazzati}(1999)}]{GhiselliniLazzati1999}
{Ghisellini}, G., \& {Lazzati}, D. 1999,
  \href{http://dx.doi.org/10.1046/j.1365-8711.1999.03025.x}{\JournalTitle{\mnras},
  309, L7}

\bibitem[{{Goldstein}(2010)}]{gcn11403}
{Goldstein}, A. 2010, \JournalTitle{GRB Coordinates Network}, 11403

\bibitem[{{Granot} {et~al.}(2015){Granot}, {Piran}, {Bromberg}, {Racusin}, \&
  {Daigne}}]{granot}
{Granot}, J., {Piran}, T., {Bromberg}, O., {Racusin}, J.~L., \& {Daigne}, F.
  2015, \href{http://dx.doi.org/10.1007/s11214-015-0191-6}{\JournalTitle{\ssr},
  191, 471}

\bibitem[{{Guidorzi} \& {Melandri}(2012)}]{gcn13092}
{Guidorzi}, C., \& {Melandri}, A. 2012, \JournalTitle{GRB Coordinates Network},
  13092

\bibitem[{{Guidorzi} {et~al.}(2011){Guidorzi}, {Melandri}, {Steele}, {Mundell},
  {Kobayashi}, \& {Gomboc}}]{gcn11537}
{Guidorzi}, C., {Melandri}, A., {Steele}, I.~A., {et~al.} 2011,
  \JournalTitle{GRB Coordinates Network}, 11537

\bibitem[{{Guidorzi} \& {Mundell}(2012)}]{gcn13651}
{Guidorzi}, C., \& {Mundell}, C.~G. 2012, \JournalTitle{GRB Coordinates
  Network}, 13651

\bibitem[{{Guidorzi} {et~al.}(2010){Guidorzi}, {Smith}, {Mundell}, {Steele},
  {Kobayashi}, \& {Gomboc}}]{gcn11397}
{Guidorzi}, C., {Smith}, R.~J., {Mundell}, C.~G., {et~al.} 2010,
  \JournalTitle{GRB Coordinates Network}, 11397

\bibitem[{{Guidorzi} {et~al.}(2006){Guidorzi}, {Monfardini}, {Gomboc},
  {Mottram}, {Mundell}, {Steele}, {Carter}, {Bode}, {Smith}, {Fraser},
  {Burgdorf}, \& {Newsam}}]{pasp-paper}
{Guidorzi}, C., {Monfardini}, A., {Gomboc}, A., {et~al.} 2006,
  \href{http://dx.doi.org/10.1086/499289}{\JournalTitle{\pasp}, 118, 288}

\bibitem[{{Holland} \& {Hoversten}(2010)}]{gcn11062}
{Holland}, S.~T., \& {Hoversten}, E.~A. 2010, \JournalTitle{GRB Coordinates
  Network}, 11062

\bibitem[{{Japelj} {et~al.}(2014){Japelj}, {Kopa{\v c}}, {Kobayashi},
  {Harrison}, {Guidorzi}, {Virgili}, {Mundell}, {Melandri}, \&
  {Gomboc}}]{japelj2014}
{Japelj}, J., {Kopa{\v c}}, D., {Kobayashi}, S., {et~al.} 2014,
  \href{http://dx.doi.org/10.1088/0004-637X/785/2/84}{\JournalTitle{\apj}, 785,
  84}

\bibitem[{{Jermak} {et~al.}(2016){Jermak}, {Steele}, {Lindfors}, {Hovatta},
  {Nilsson}, {Lamb}, {Mundell}, {Barres de Almeida}, {Berdyugin}, {Kadenius},
  {Reinthal}, \& {Takalo}}]{jermak-mnras}
{Jermak}, H., {Steele}, I.~A., {Lindfors}, E., {et~al.} 2016,
  \href{http://dx.doi.org/10.1093/mnras/stw1770}{\JournalTitle{\mnras}, 462,
  4267}

\bibitem[{{Jermak}(2016)}]{jermak-phd}
{Jermak}, H.~E. 2016, \JournalTitle{PhD Thesis, Liverpool JMU}

\bibitem[{{Jordi} {et~al.}(2006){Jordi}, {Grebel}, \& {Ammon}}]{jordi}
{Jordi}, K., {Grebel}, E.~K., \& {Ammon}, K. 2006,
  \href{http://dx.doi.org/10.1051/0004-6361:20066082}{\JournalTitle{\aap}, 460,
  339}

\bibitem[{{Klotz} {et~al.}(2011){Klotz}, {Gendre}, {Boer}, \&
  {Atteia}}]{gcn12022}
{Klotz}, A., {Gendre}, B., {Boer}, M., \& {Atteia}, J.~L. 2011,
  \JournalTitle{GRB Coordinates Network}, 12022

\bibitem[{{Klotz} {et~al.}(2012){Klotz}, {Gendre}, {Boer}, \&
  {Atteia}}]{gcn13290}
---. 2012, \JournalTitle{GRB Coordinates Network}, 13290

\bibitem[{{Kobayashi}(2000)}]{kobayashi2000}
{Kobayashi}, S. 2000,
  \href{http://dx.doi.org/10.1086/317869}{\JournalTitle{\apj}, 545, 807}

\bibitem[{{Kobayashi} {et~al.}(1999){Kobayashi}, {Piran}, \& {Sari}}]{k99}
{Kobayashi}, S., {Piran}, T., \& {Sari}, R. 1999,
  \href{http://dx.doi.org/10.1086/306868}{\JournalTitle{\apj}, 513, 669}

\bibitem[{{Komissarov} {et~al.}(2009){Komissarov}, {Vlahakis}, {K{\"o}nigl}, \&
  {Barkov}}]{Komissarov}
{Komissarov}, S.~S., {Vlahakis}, N., {K{\"o}nigl}, A., \& {Barkov}, M.~V. 2009,
  \href{http://dx.doi.org/10.1111/j.1365-2966.2009.14410.x}{\JournalTitle{\mnras},
  394, 1182}

\bibitem[{{Kopa{\v c}} {et~al.}(2013){Kopa{\v c}}, {Kobayashi}, {Gomboc},
  {Japelj}, {Mundell}, {Guidorzi}, {Melandri}, {Bersier}, {Cano}, {Smith},
  {Steele}, \& {Virgili}}]{kopac2013}
{Kopa{\v c}}, D., {Kobayashi}, S., {Gomboc}, A., {et~al.} 2013,
  \href{http://dx.doi.org/10.1088/0004-637X/772/1/73}{\JournalTitle{\apj}, 772,
  73}

\bibitem[{{Kopa{\v c}} {et~al.}(2015){Kopa{\v c}}, {Mundell}, {Japelj},
  {Arnold}, {Steele}, {Guidorzi}, {Dichiara}, {Kobayashi}, {Gomboc},
  {Harrison}, {Lamb}, {Melandri}, {Smith}, {Virgili}, {Castro-Tirado},
  {Gorosabel}, {J{\"a}rvinen}, {S{\'a}nchez-Ram{\'{\i}}rez}, {Oates}, \&
  {Jel{\'{\i}}nek}}]{kopac15}
{Kopa{\v c}}, D., {Mundell}, C.~G., {Japelj}, J., {et~al.} 2015,
  \href{http://dx.doi.org/10.1088/0004-637X/813/1/1}{\JournalTitle{\apj}, 813,
  1}

\bibitem[{{Kuroda} {et~al.}(2012){Kuroda}, {Hanayama}, {Miyaji}, {Watanabe},
  {Yanagisawa}, {Nagayama}, {Yoshida}, {Ohta}, \& {Kawai}}]{gcn13465}
{Kuroda}, D., {Hanayama}, H., {Miyaji}, T., {et~al.} 2012, \JournalTitle{GRB
  Coordinates Network}, 13465

\bibitem[{{Lacluyze} {et~al.}(2011){Lacluyze}, {Maturi}, {Ivarsen}, {Haislip},
  {Reichart}, {Moore}, {Trotter}, {Foster}, {Egger}, {Oza}, {Speckhard},
  {Crain}, \& {Nysewander}}]{gcn12024}
{Lacluyze}, A., {Maturi}, M., {Ivarsen}, K., {et~al.} 2011, \JournalTitle{GRB
  Coordinates Network}, 12024

\bibitem[{{Lazzati} {et~al.}(2004){Lazzati}, {Rossi}, {Ghisellini}, \&
  {Rees}}]{Lazzati2004}
{Lazzati}, D., {Rossi}, E., {Ghisellini}, G., \& {Rees}, M.~J. 2004,
  \href{http://dx.doi.org/10.1111/j.1365-2966.2004.07387.x}{\JournalTitle{\mnras},
  347, L1}

\bibitem[{{Lien} {et~al.}(2016){Lien}, {Sakamoto}, {Barthelmy}, {Baumgartner},
  {Cannizzo}, {Chen}, {Collins}, {Cummings}, {Gehrels}, {Krimm}, {Markwardt},
  {Palmer}, {Stamatikos}, {Troja}, \& {Ukwatta}}]{lien16}
{Lien}, A., {Sakamoto}, T., {Barthelmy}, S.~D., {et~al.} 2016,
  \href{http://dx.doi.org/10.3847/0004-637X/829/1/7}{\JournalTitle{ApJ}, 829,
  7}

\bibitem[{{Lin} {et~al.}(2017){Lin}, {Li}, \& {Chang}}]{Lin2017}
{Lin}, H.-N., {Li}, X., \& {Chang}, Z. 2017,
  \href{http://dx.doi.org/10.1088/1674-1137/41/4/045101}{\JournalTitle{Chinese
  Physics C}, 41, 045101}

\bibitem[{{Medvedev} \& {Loeb}(1999)}]{med99}
{Medvedev}, M.~V., \& {Loeb}, A. 1999,
  \href{http://dx.doi.org/10.1086/308038}{\JournalTitle{\apj}, 526, 697}

\bibitem[{{Melandri} {et~al.}(2014){Melandri}, {Virgili}, {Guidorzi},
  {Bernardini}, {Kobayashi}, {Mundell}, {Gomboc}, {Dintinjana}, {Hentunen},
  {Japelj}, {Kopa{\v c}}, {Kuroda}, {Morgan}, {Steele}, {Quadri}, {Arici},
  {Arnold}, {Girelli}, {Hanayama}, {Kawai}, {Miku{\v z}}, {Nissinen}, {Salmi},
  {Smith}, {Strabla}, {Tonincelli}, \& {Quadri}}]{melandri}
{Melandri}, A., {Virgili}, F.~J., {Guidorzi}, C., {et~al.} 2014,
  \href{http://dx.doi.org/10.1051/0004-6361/201424338}{\JournalTitle{\aap},
  572, A55}

\bibitem[{{Morgan} {et~al.}(2014){Morgan}, {Perley}, {Cenko}, {Bloom},
  {Cucchiara}, {Richards}, {Filippenko}, {Haislip}, {LaCluyze}, {Corsi},
  {Melandri}, {Cobb}, {Gomboc}, {Horesh}, {James}, {Li}, {Mundell}, {Reichart},
  \& {Steele}}]{morgan}
{Morgan}, A.~N., {Perley}, D.~A., {Cenko}, S.~B., {et~al.} 2014,
  \href{http://dx.doi.org/10.1093/mnras/stu344}{\JournalTitle{\mnras}, 440,
  1810}

\bibitem[{{Mundell} {et~al.}(2011){Mundell}, {Melandri}, \&
  {Tanvir}}]{gcn11858}
{Mundell}, C.~G., {Melandri}, A., \& {Tanvir}, N. 2011, \JournalTitle{GRB
  Coordinates Network}, 11858

\bibitem[{{Mundell} {et~al.}(2013){Mundell}, {Kopa{\v c}}, {Arnold}, {Steele},
  {Gomboc}, {Kobayashi}, {Harrison}, {Smith}, {Guidorzi}, {Virgili},
  {Melandri}, \& {Japelj}}]{mundell-nature}
{Mundell}, C.~G., {Kopa{\v c}}, D., {Arnold}, D.~M., {et~al.} 2013,
  \href{http://dx.doi.org/10.1038/nature12814}{\JournalTitle{\nat}, 504, 119}

\bibitem[{{Naghizadeh-Khouei} \& {Clarke}(1993)}]{nag93}
{Naghizadeh-Khouei}, J., \& {Clarke}, D. 1993, \JournalTitle{\aap}, 274, 968

\bibitem[{{Nava} {et~al.}(2016){Nava}, {Nakar}, \& {Piran}}]{Nava2016}
{Nava}, L., {Nakar}, E., \& {Piran}, T. 2016,
  \href{http://dx.doi.org/10.1093/mnras/stv2434}{\JournalTitle{\mnras}, 455,
  1594}

\bibitem[{{Nousek} {et~al.}(2006){Nousek}, {Kouveliotou}, {Grupe}, {Page},
  {Granot}, {Ramirez-Ruiz}, {Patel}, {Burrows}, {Mangano}, {Barthelmy},
  {Beardmore}, {Campana}, {Capalbi}, {Chincarini}, {Cusumano}, {Falcone},
  {Gehrels}, {Giommi}, {Goad}, {Godet}, {Hurkett}, {Kennea}, {Moretti},
  {O'Brien}, {Osborne}, {Romano}, {Tagliaferri}, \& {Wells}}]{nousek2006}
{Nousek}, J.~A., {Kouveliotou}, C., {Grupe}, D., {et~al.} 2006,
  \href{http://dx.doi.org/10.1086/500724}{\JournalTitle{\apj}, 642, 389}

\bibitem[{{Papoulis}(1984)}]{rayleigh}
{Papoulis}, A. 1984, {Probability, random variables and stochastic processes}

\bibitem[{{Perley} \& {Tanvir}(2012)}]{gcn13133}
{Perley}, D.~A., \& {Tanvir}, N.~R. 2012, \JournalTitle{GRB Coordinates
  Network}, 13133

\bibitem[{{Piran}(1999)}]{piran1999}
{Piran}, T. 1999,
  \href{http://dx.doi.org/10.1016/S0370-1573(98)00127-6}{\JournalTitle{\physrep},
  314, 575}

\bibitem[{{Rossi} {et~al.}(2004){Rossi}, {Lazzati}, {Salmonson}, \&
  {Ghisellini}}]{Rossi2004}
{Rossi}, E.~M., {Lazzati}, D., {Salmonson}, J.~D., \& {Ghisellini}, G. 2004,
  \href{http://dx.doi.org/10.1111/j.1365-2966.2004.08165.x}{\JournalTitle{\mnras},
  354, 86}

\bibitem[{{Sari}(1999)}]{Sari1999}
{Sari}, R. 1999, \href{http://dx.doi.org/10.1086/312294}{\JournalTitle{\apjl},
  524, L43}

\bibitem[{{Sari} \& {Piran}(1995)}]{saripiran95}
{Sari}, R., \& {Piran}, T. 1995,
  \href{http://dx.doi.org/10.1086/309835}{\JournalTitle{\apjl}, 455, L143}

\bibitem[{{Sari} \& {Piran}(1999)}]{SariPiran1999}
---. 1999, \href{http://dx.doi.org/10.1086/312039}{\JournalTitle{\apjl}, 517,
  L109}

\bibitem[{{Sari} {et~al.}(1998){Sari}, {Piran}, \& {Narayan}}]{Sari1998}
{Sari}, R., {Piran}, T., \& {Narayan}, R. 1998,
  \href{http://dx.doi.org/10.1086/311269}{\JournalTitle{\apjl}, 497, L17}

\bibitem[{{Schlafly} \& {Finkbeiner}(2011)}]{extinction}
{Schlafly}, E.~F., \& {Finkbeiner}, D.~P. 2011,
  \href{http://dx.doi.org/10.1088/0004-637X/737/2/103}{\JournalTitle{\apj},
  737, 103}

\bibitem[{{Schmidt} {et~al.}(1992){Schmidt}, {Elston}, \& {Lupie}}]{schmidt}
{Schmidt}, G.~D., {Elston}, R., \& {Lupie}, O.~L. 1992,
  \href{http://dx.doi.org/10.1086/116341}{\JournalTitle{\aj}, 104, 1563}

\bibitem[{{Simmons} \& {Stewart}(1985)}]{simmons}
{Simmons}, J.~F.~L., \& {Stewart}, B.~G. 1985, \JournalTitle{\aap}, 142, 100

\bibitem[{{Smith} {et~al.}(2002){Smith}, {Tucker}, {Kent}, {Richmond},
  {Fukugita}, {Ichikawa}, {Ichikawa}, {Jorgensen}, {Uomoto}, {Gunn}, {Hamabe},
  {Watanabe}, {Tolea}, {Henden}, {Annis}, {Pier}, {McKay}, {Brinkmann}, {Chen},
  {Holtzman}, {Shimasaku}, \& {York}}]{sloan}
{Smith}, J.~A., {Tucker}, D.~L., {Kent}, S., {et~al.} 2002,
  \href{http://dx.doi.org/10.1086/339311}{\JournalTitle{\aj}, 123, 2121}

\bibitem[{{Steele}(2001)}]{ratcam}
{Steele}, I.~A. 2001,
  \href{http://dx.doi.org/10.1002/1521-3994(200112)322:5/6<307::AID-ASNA307>3.0.CO;2-2}{\JournalTitle{Astronomische
  Nachrichten}, 322, 307}

\bibitem[{{Steele} {et~al.}(2010){Steele}, {Bates}, {Guidorzi}, {Mottram},
  {Mundell}, \& {Smith}}]{steele2010}
{Steele}, I.~A., {Bates}, S.~D., {Guidorzi}, C., {et~al.} 2010,
  \href{http://dx.doi.org/10.1117/12.856842}{in \procspie, Vol. 7735,
  Ground-based and Airborne Instrumentation for Astronomy III}, 773549

\bibitem[{{Steele} {et~al.}(2014){Steele}, {Mottram}, {Smith}, \&
  {Barnsley}}]{steele14}
{Steele}, I.~A., {Mottram}, C.~J., {Smith}, R.~J., \& {Barnsley}, R.~M. 2014,
  \href{http://dx.doi.org/10.1117/12.2056602}{in \procspie, Vol. 9154, High
  Energy, Optical, and Infrared Detectors for Astronomy VI}, 915428

\bibitem[{{Steele} {et~al.}(2009){Steele}, {Mundell}, {Smith}, {Kobayashi}, \&
  {Guidorzi}}]{SteeleNature2009}
{Steele}, I.~A., {Mundell}, C.~G., {Smith}, R.~J., {Kobayashi}, S., \&
  {Guidorzi}, C. 2009,
  \href{http://dx.doi.org/10.1038/nature08590}{\JournalTitle{\nat}, 462, 767}

\bibitem[{{Steele} {et~al.}(2004){Steele}, {Smith}, {Rees}, {Baker}, {Bates},
  {Bode}, {Bowman}, {Carter}, {Etherton}, {Ford}, {Fraser}, {Gomboc}, {Lett},
  {Mansfield}, {Marchant}, {Medrano-Cerda}, {Mottram}, {Raback}, {Scott},
  {Tomlinson}, \& {Zamanov}}]{steele04}
{Steele}, I.~A., {Smith}, R.~J., {Rees}, P.~C., {et~al.} 2004,
  \href{http://dx.doi.org/10.1117/12.551456}{in \procspie, Vol. 5489,
  Ground-based Telescopes, ed. J.~M. {Oschmann}, Jr.}, 679

\bibitem[{{Steele} {et~al.}(2006){Steele}, {Bates}, {Carter}, {Clarke},
  {Gomboc}, {Guidorzi}, {Melandri}, {Monfardini}, {Mottram}, {Mundell},
  {Scott}, {Smith}, \& {Swindlehurst}}]{ringo}
{Steele}, I.~A., {Bates}, S.~D., {Carter}, D., {et~al.} 2006,
  \href{http://dx.doi.org/10.1117/12.670756}{in \procspie, Vol. 6269, Society
  of Photo-Optical Instrumentation Engineers (SPIE) Conference Series}, 62695M

\bibitem[{{Tello} {et~al.}(2012){Tello}, {Sanchez-Ramirez}, {Gorosabel},
  {Castro-Tirado}, {Rivero}, {Gomez-Velarde}, \& {Klotz}}]{gcn13118}
{Tello}, J.~C., {Sanchez-Ramirez}, R., {Gorosabel}, J., {et~al.} 2012,
  \JournalTitle{GRB Coordinates Network}, 13118

\bibitem[{{Turnshek} {et~al.}(1990){Turnshek}, {Bohlin}, {Williamson}, {Lupie},
  {Koornneef}, \& {Morgan}}]{turnshek}
{Turnshek}, D.~A., {Bohlin}, R.~C., {Williamson}, II, R.~L., {et~al.} 1990,
  \href{http://dx.doi.org/10.1086/115413}{\JournalTitle{\aj}, 99, 1243}

\bibitem[{{Uehara} {et~al.}(2012){Uehara}, {Toma}, {Kawabata}, {Chiyonobu},
  {Fukazawa}, {Ikejiri}, {Inoue}, {Itoh}, {Komatsu}, {Miyamoto}, {Mizuno},
  {Nagae}, {Nakaya}, {Ohsugi}, {Sakimoto}, {Sasada}, {Tanaka}, {Uemura},
  {Yamanaka}, {Yamashita}, {Yamazaki}, \& {Yoshida}}]{uehara2012}
{Uehara}, T., {Toma}, K., {Kawabata}, K.~S., {et~al.} 2012,
  \href{http://dx.doi.org/10.1088/2041-8205/752/1/L6}{\JournalTitle{\apjl},
  752, L6}

\bibitem[{{Virgili} {et~al.}(2012){Virgili}, {Guidorzi}, {Melandri}, \&
  {Mundell}}]{gcn13006}
{Virgili}, F.~J., {Guidorzi}, C., {Melandri}, A., \& {Mundell}, C.~G. 2012,
  \JournalTitle{GRB Coordinates Network}, 13006

\bibitem[{{Wiersema} {et~al.}(2014){Wiersema}, {Covino}, {Toma}, {van der
  Horst}, {Varela}, {Min}, {Greiner}, {Starling}, {Tanvir}, {Wijers},
  {Campana}, {Curran}, {Fan}, {Fynbo}, {Gorosabel}, {Gomboc}, {G{\"o}tz},
  {Hjorth}, {Jin}, {Kobayashi}, {Kouveliotou}, {Mundell}, {O'Brien}, {Pian},
  {Rowlinson}, {Russell}, {Salvaterra}, {di Serego Alighieri}, {Tagliaferri},
  {Vergani}, {Elliott}, {Fari{\~n}a}, {Hartoog}, {Karjalainen}, {Klose},
  {Knust}, {Levan}, {Schady}, {Sudilovsky}, \& {Willingale}}]{Wiersema2014}
{Wiersema}, K., {Covino}, S., {Toma}, K., {et~al.} 2014,
  \href{http://dx.doi.org/10.1038/nature13237}{\JournalTitle{\nat}, 509, 201}

\bibitem[{{Wijers} {et~al.}(1999){Wijers}, {Vreeswijk}, {Galama}, {Rol}, {van
  Paradijs}, {Kouveliotou}, {Giblin}, {Masetti}, {Palazzi}, {Pian}, {Frontera},
  {Nicastro}, {Falomo}, {Soffitta}, \& {Piro}}]{Wijers1999}
{Wijers}, R.~A.~M.~J., {Vreeswijk}, P.~M., {Galama}, T.~J., {et~al.} 1999,
  \href{http://dx.doi.org/10.1086/312262}{\JournalTitle{\apjl}, 523, L33}

\bibitem[{{Yonetoku} {et~al.}(2011){Yonetoku}, {Murakami}, {Gunji}, {Mihara},
  {Toma}, {Sakashita}, {Morihara}, {Takahashi}, {Toukairin}, {Fujimoto},
  {Kodama}, {Kubo}, \& {IKAROS Demonstration Team}}]{Yonetoku}
{Yonetoku}, D., {Murakami}, T., {Gunji}, S., {et~al.} 2011,
  \href{http://dx.doi.org/10.1088/2041-8205/743/2/L30}{\JournalTitle{\apjl},
  743, L30}

\bibitem[{{Zhang} {et~al.}(2006){Zhang}, {Fan}, {Dyks}, {Kobayashi},
  {M{\'e}sz{\'a}ros}, {Burrows}, {Nousek}, \& {Gehrels}}]{zhang2006}
{Zhang}, B., {Fan}, Y.~Z., {Dyks}, J., {et~al.} 2006,
  \href{http://dx.doi.org/10.1086/500723}{\JournalTitle{\apj}, 642, 354}

\bibitem[{{Zhang} {et~al.}(2003){Zhang}, {Kobayashi}, \&
  {M{\'e}sz{\'a}ros}}]{zhang2003}
{Zhang}, B., {Kobayashi}, S., \& {M{\'e}sz{\'a}ros}, P. 2003,
  \href{http://dx.doi.org/10.1086/377363}{\JournalTitle{\apj}, 595, 950}

\bibitem[{{Zhang} \& {M{\'e}sz{\'a}ros}(2004)}]{zhang2004}
{Zhang}, B., \& {M{\'e}sz{\'a}ros}, P. 2004,
  \href{http://dx.doi.org/10.1142/S0217751X0401746X}{\JournalTitle{International
  Journal of Modern Physics A}, 19, 2385}

\bibitem[{{Zhang} \& {Yan}(2011)}]{ZhangYan2011}
{Zhang}, B., \& {Yan}, H. 2011,
  \href{http://dx.doi.org/10.1088/0004-637X/726/2/90}{\JournalTitle{\apj}, 726,
  90}

\bibitem[{{Zheng} {et~al.}(2012){Zheng}, {Shen}, {Sakamoto}, {Beardmore}, {De
  Pasquale}, {Wu}, {Gorosabel}, {Urata}, {Sugita}, {Zhang}, {Pozanenko},
  {Nissinen}, {Sahu}, {Im}, {Ukwatta}, {Andreev}, {Klunko}, {Volnova},
  {Akerlof}, {Anto}, {Barthelmy}, {Breeveld}, {Carsenty},
  {Castillo-Carri{\'o}n}, {Castro-Tirado}, {Chester}, {Chuang}, {Cunniffe}, {De
  Ugarte Postigo}, {Duffard}, {Flewelling}, {Gehrels}, {G{\"u}ver}, {Guziy},
  {Hentunen}, {Huang}, {Jel{\'{\i}}nek}, {Koch}, {Kub{\'a}nek}, {Kuin},
  {McKay}, {Mottola}, {Oates}, {O'Brien}, {Ohno}, {Page}, {Pandey}, {P{\'e}rez
  del Pulgar}, {Rujopakarn}, {Rykoff}, {Salmi}, {S{\'a}nchez-Ram{\'{\i}}rez},
  {Schaefer}, {Sergeev}, {Sonbas}, {Sota}, {Tello}, {Yamaoka}, {Yost}, \&
  {Yuan}}]{zheng2012}
{Zheng}, W., {Shen}, R.~F., {Sakamoto}, T., {et~al.} 2012,
  \href{http://dx.doi.org/10.1088/0004-637X/751/2/90}{\JournalTitle{\apj}, 751,
  90}

\end{thebibliography}

\end{document}